\documentclass[journal,final]{IEEEtran}
\usepackage[noadjust]{cite}
\usepackage{amsmath}  \interdisplaylinepenalty=2500
\usepackage{amssymb}
\usepackage{array}
\usepackage{multirow}
\usepackage{mathtools}
\usepackage{subcaption}

\usepackage{ifpdf}
\ifpdf
\usepackage[pdftex]{graphicx} 
\usepackage{epstopdf}         
\else
\usepackage{graphicx}         
\fi

\begin{document}

\title{Timing, Carrier, and Frame Synchronization of\\ Burst-Mode CPM}

\author{Ehsan~Hosseini,~\IEEEmembership{Student Member,~IEEE,}
        and~Erik~Perrins,~\IEEEmembership{Senior Member,~IEEE}
\thanks{This paper will be presented in part at the IEEE Global Telecommunications Conference, Atlanta, Georgia, USA, December 2013.}        
\thanks{E. Hosseini and E. Perrins are with the Department of Electrical Engineering and Computer Science,
University of Kansas, Lawrence, KS 66045 USA (\hbox{e-mail}:~ehsan@ku.edu, esp@ieee.org).}
\thanks{}        
\thanks{Copyright 2013 IEEE. Personal use of this material is permitted. However, permission to reprint/republish this material for advertising or promotional purposes or for creating new collective works for resale or redistribution to servers or lists, or to reuse any copyrighted component of this work in other works must be obtained from the IEEE.}
\thanks{Published in: Ehsan Hosseini and Erik Perrins, ``Timing, Carrier, and Frame Synchronization of Burst-Mode CPM,'' \textit{IEEE Transactions on Communications}, vol.61, no.12, pp.5125-5138, December 2013}
\thanks{DOI: 10.1109/TCOMM.2013.111613.130667}
\thanks{URL: http://ieeexplore.ieee.org/stamp/stamp.jsp?tp=\&arnumber= 6678035\&isnumber=6689285}
}

\maketitle

\begin{abstract}
In this paper, we propose a complete synchronization algorithm for continuous phase modulation (CPM) signals in burst-mode transmission over additive white Gaussian noise (AWGN) channels. The timing and carrier recovery are performed through a data-aided (DA) maximum likelihood algorithm, which jointly estimates symbol timing, carrier phase, and frequency offsets based on an optimized synchronization preamble. Our algorithm estimates the frequency offset via a one-dimensional grid search, after which symbol timing and carrier phase are computed via simple closed-form expressions. The mean-square error (MSE) of the algorithm's estimates reveals that it performs very close to the theoretical Cram\'{e}r-Rao bound (CRB) for various CPMs at signal-to-noise ratios (SNRs) as low as 0 dB. Furthermore, we present a frame synchronization algorithm that detects the arrival of bursts and estimates the start-of-signal. We simulate the performance of the frame synchronization algorithm along with the timing and carrier recovery algorithm. The bit error rate results demonstrate near ideal synchronization performance for low SNRs and short preambles.     
\end{abstract}
\section{Introduction}
\PARstart{C}{ontinuous} phase modulation (CPM) \cite{Aulin1981} is a highly bandwidth and power-efficient digital transmission scheme, which allows designers to employ non-linear power amplifiers. It has been an attractive choice for time-division multiple-access (TDMA) networks where data or voice is transmitted in a burst-mode fashion. Examples of such systems are the well-known GSM cellular standard \cite{Gsm2} and the next generation aeronautical telemetry standard \cite{Geoghegan2011}. Despite the attractive features of CPM, the receiver complexity is high due to the inherent memory of the modulation, and it requires maximum likelihood sequence detection (MLSD) for the best performance \cite{Anderson1986}.
\par 
Another source of receiver complexity is the synchronization task, especially in burst-mode transmissions where the \textit{warm-up} or \textit{acquisition} time must be kept as small as possible. This task has become even more challenging due to the introduction of powerful error correction codes such as low-density parity check (LDPC) codes, which require accurate synchronization at low signal-to-noise ratios (SNRs) in order to achieve the full coding gain. \textit{Feedforward} synchronization is a common approach in this type of application since it requires a shorter acquisition time compared to closed-loop methods \cite{Mengali1997}. Moreover, a known synchronization \textit{preamble} is usually appended to the beginning of each burst, which assists the synchronization via data-aided (DA) algorithms.
\par 
The majority of works on synchronization of CPM in burst-mode transmissions have addressed minimum-shift keying (MSK)-type modulations using non-data-aided (NDA) algorithms, e.g. \cite{260769,Morelli1998,Morelli2001}. In addition to their limited application, these methods do not perform as well as DA algorithms in low SNRs. Huang \textit{et al.} \cite{Huang2000} have proposed a feedforward DA joint symbol timing and frequency offset estimation algorithm for Gaussian MSK (GMSK) signals. The performance of this \textit{ad-hoc} method relies on the amount of frequency offset and sample timing error. A feedforward NDA symbol timing estimation is presented in \cite{D'Andrea1996}, which can work for general CPMs in principle. However, its performance degrades in case of partial response schemes. A few DA synchronization algorithms have been presented in the literature for general CPM signals in different environments \cite{Huber1992,Tang2001,Zhao2006a,Shen2007}. Huber and Liu \cite{Huber1992} proposed a maximum likelihood (ML) joint timing and phase synchronization algorithm for additive white Gaussian noise (AWGN) channels. In a related work \cite{Tang2001}, the Walsh transform is used in order to derive the synchronization algorithm. Both of these algorithms assume the timing offset is much smaller than the symbol duration in order to function properly. This limits their application in burst-mode feedforward receivers as the timing offset in practice can have any arbitrary value. Another DA joint phase and timing estimation algorithm is proposed in \cite{Zhao2006a}, which is based on the minimum mean-square error (MMSE) and Kalman filter criteria. Despite its robustness in time-variant channels and short preambles, this method is implemented in a closed-loop manner which requires multiple initialization steps. Moreover, its mean-square error (MSE) is shown to be significantly larger than the Cram\'{e}r-Rao bound (CRB) even at high SNRs. Another DA algorithm is proposed in \cite{Shen2007} for space-time coded CPM over Rayleigh channels, which only tackles the symbol timing estimation. One important issue with all the aforementioned DA algorithms is that the carrier frequency offset has not been taken into account. Blind frequency estimators such as \cite{D'Andrea1995,Bianchi2005} can be employed prior to symbol timing and phase estimations. However, the accuracy of these frequency estimators is far above the CRB \cite{D'Andrea1995} especially in low to moderate SNRs \cite{Bianchi2005}. Residual frequency offsets result in poor timing and phase estimators as well as signal demodulation.   
\par
Another challenge in synchronization of burst-mode signals is estimation of the burst start point, i.e. start-of-signal (SoS). This task, which will be referred to as \textit{frame synchronization}, is crucial in DA algorithms where the boundaries of the known preamble have to be identified. Several sophisticated frame synchronization algorithms \cite{Gansman1997,Lee2002a,Pedone2010} have been proposed for phase shift-keying (PSK) signals in AWGN where frequency offset is present. The performance of the algorithm in \cite{Gansman1997} depends on the amount of frequency offset, which has to be much smaller than the symbol rate. Choi and Lee \cite{Lee2002a} have assumed continuous transmissions, where the preamble is surrounded by random data. Although burst-mode transmission is introduced in \cite{Pedone2010}, the authors have assumed there is no guard interval between bursts and the preamble is preceded by random data (similar to the continuous mode). Moreover, it assumes the tentative location of the preamble is known within an uncertainty window. Such a knowledge might not be always available particularly when the receiver is just powered on.
\par 
In this work, we present a feedforward DA ML algorithm for joint estimation of frequency offset, symbol timing, and carrier phase in burst-mode CPM signals. The proposed approach takes advantage of the optimized preamble of \cite{Hosseini2013}, which jointly minimizes the CRBs for all three synchronization parameters. We show that the proposed algorithm is capable of performing quite close to the CRB for various CPMs and SNRs. Although we consider an AWGN channel, the results can be applied to time-varying channels too since practical wireless channels can be assumed to be static during the preamble period. In such environments, the estimation results should be used in conjunction with tracking algorithms such as \cite{Morelli1997}. Additionally, we present a frame synchronization algorithm that detects the arrival of bursts and estimates the SoS within ideally one sample time. We discuss how our approach extends the frame synchronization algorithms in \cite{Gansman1997,Lee2002a} to our problem, i.e. CPM signals and burst-mode transmissions. We note that the order in which these two problems are addressed in this paper is the reverse of their implementation in practice where frame synchronization must be applied prior to timing and carrier recovery.
\par
The remainder of this paper is organized as follows. Section II introduces our burst-mode transmission model. In Section III, the joint ML timing and carrier estimation is proposed. Section IV describes the frame synchronization algorithm. Simulation results of the synchronization algorithm are reported in Section V, and Section VI concludes the paper.      
\section{Burst-Mode Transmission Model}
\begin{figure}[t]
\centering
\includegraphics[width=0.9\linewidth]{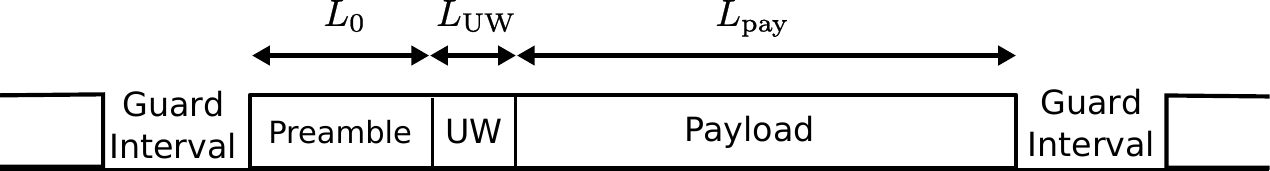}
\caption{The Burst-Mode Transmission Model.}
\label{fig:BurstStruc}
\end{figure}

In our model, we consider transmission of disjoint packets of data, i.e. bursts. The transmitter is assumed to be turned on at an unknown time in order to transmit a single burst after which it is turned off again. Each burst has a known duration and structure at the receiver, which is depicted in Fig. \ref{fig:BurstStruc} and consists of three parts. The first part is the synchronization preamble or training sequence. It consists of $L_0$ known and optimized data symbols which are used to estimate synchronization parameters. Although the preamble can be used for channel estimation too, we only focus on the synchronization task. The next section in the burst is denoted as the \textit{unique word} (UW), which is utilized to identify the bursts and determine the location of data symbols within a burst. It is assumed to be a pseudo-random sequence of $L_{\text{UW}}$ symbols. The last part is the \textit{payload}, which carries $L_{\text{pay}}$ information symbols.
\par  
We consider CPM signaling for transmission of bursts in our model. The complex baseband CPM signal during transmission of each burst can be expressed as 
\begin{equation}
s(t) = \sqrt{\frac{E_s}{T_s}}\exp\{j\phi(t;\boldsymbol{\alpha})\}
\end{equation}
where $E_s$ is the energy per transmitted symbol. The phase of the signal $\phi(t;\boldsymbol{\alpha})$ is represented as
\begin{equation}
\phi(t;\boldsymbol{\alpha}) = 2\pi h \sum_{i=0}^{L_b-1} \alpha_i q(t-iT_s)
\end{equation}
where $\alpha_i$ is the sequence of $M$-ary data symbols selected from the set of $\{\pm 1, \pm 3, \dots , \pm (M-1)\}$. $L_b$ is the total number of such symbols in a burst, that is $L_b = L_0+L_{\text{uw}}+L_{\text{pay}}$. The variable $h$ is the modulation index, which can vary from symbol to symbol in the case of \textit{multi-h} CPM. The waveform $q(t)$ is the \textit{phase response} of CPM and in general is represented as the integral of the \textit{frequency pulse} $g(t)$ with a duration of $LT_s$. If $L=1$, the signal is called \textit{full response} CPM, and for $L>1$, it is called \textit{partial response} CPM. In CPM literature, there are two well-known frequency pulses denoted by length $L$ rectangular ($L$REC) and length $L$ raised cosine ($L$RC) \cite{Proakis}. Another commonly-used frequency pulse is the Gaussian minimum-shift keying (GMSK) pulse with bandwidth parameter $BT_s$. In our discussion, we will use $BT_s=0.3$, which is the bandwidth parameter in the GSM standard.   
\par 
Assuming transmission over an AWGN channel, the complex baseband representation of the received signal is
\begin{equation}
r(t) = e^{j(2\pi f_d t + \theta)} s(t-\tau) + w(t)
\label{eq:cpm-received}
\end{equation} 
where $\theta$ is the unknown carrier phase, $f_d$ is the frequency offset, $\tau$ is the timing offset, and $w(t)$ is complex baseband AWGN with zero mean and power spectral density $N_0$. We denote the transmitted data symbols during the preamble by $\boldsymbol{\alpha} = [\alpha_0,\alpha_1,\cdots,\alpha_{L_0-1}]$. Our goal is to determine the synchronization parameters, i.e. $\mathbf{u} = [f_d,\theta,\tau]^T$, by observing the preamble portion of the burst, which corresponds to $\boldsymbol{\alpha}$. Here,  it is assumed that $\mathbf u$ is a vector of unknown but deterministic parameters which are to be \textit{jointly} estimated at the receiver. Note that $\boldsymbol{\alpha}$ is implicit in the definition of $s(t)$.
\par 
Since data arrives in bursts at the receiver, $\tau$ can assume any value. However, a DA estimator requires the approximate knowledge of $\tau$ in order to perform the estimation algorithm on the received preamble. Therefore, we decompose $\tau$ into two parts based on
\begin{equation}
\tau = \mu T_s + \varepsilon T_s
\label{eq:delay}
\end{equation} 
where $\mu \geq 0$ is an integer that represents the \textit{integer delay} and $-0.5 < \varepsilon < 0.5$ represents the \textit{fractional delay}. In this work, we address these two components separately. First we assume $\mu$ is known and the goal is to estimate $\varepsilon$, $f_d$ and $\theta$. Later in Section IV, we consider estimation of $\mu$, i.e. the SoS location, regardless of $f_d$ and $\theta$ values.
\begin{figure}[t]
\centering
\includegraphics[width=0.8\linewidth]{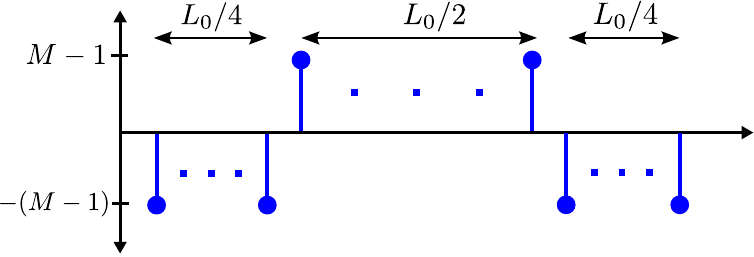}
\caption{The optimum synchronization preamble (training sequence) for $M$-ary CPM signals containing $L_0$ symbols.}
\label{fig:OptTS}
\end{figure}
\par 
The last item we need to specify is the synchronization preamble. In our recent work \cite{Hosseini2013}, we proposed the optimum training sequence for joint estimation of $\mathbf{u}$ based on the CRB criterion. This sequence, which is depicted in Fig. \ref{fig:OptTS}, minimizes the CRBs for $f_d$, $\theta$ and $\varepsilon$ simultaneously. It also has a similar pattern for the entire CPM family. We exploit the structure of the preamble in order to facilitate the algorithm design process and then to reduce its complexity.
\section{Maximum Likelihood Timing and Carrier Synchronization}
\subsection{Derivation of the Algorithm}
Reliable detection of CPM signals depends on accurate timing and carrier synchronization, which requires knowledge of $f_d$, $\theta$ and $\tau$. These parameters can be estimated via various techniques. In this work, we apply joint ML estimation in which $\boldsymbol{\alpha}$ is known to the receiver. The likelihood function for the estimation of a set of parameters from a waveform in AWGN is given in \cite{Proakis}. It can be easily shown that in our problem, i.e. when the signal is complex and constant envelope, the joint log-likelihood function (LLF) for the synchronization parameters is expressed within a constant factor of
\begin{equation}
\Lambda[r(t);\tilde{f_d},\tilde{\theta},\tilde{\varepsilon}]\! = \!\mathrm{Re} \!\left[\!
\int_{\tilde{\varepsilon} T_s}^{T_0 + \tilde{\varepsilon}T_s} \!\!e^{-j(2\pi\tilde{f_d}t+\tilde{\theta})}r(t)s^*(t-\tilde{\varepsilon}T_s)\, dt 
\right]
\label{eq:ML}
\end{equation} 
where $\tilde{f_d}$, $\tilde{\theta}$ and $\tilde{\varepsilon}$ are hypothetical values for $f_d$, $\theta$ and $\varepsilon$ respectively, and $T_0=L_0T_s$ is the preamble duration. Note that we disregard $\mu$ in this section for the sake of clarity. According to the ML criterion, we choose the trial values that maximize (\ref{eq:ML}) as the best estimates for the unknown parameters $\mathbf{u}$. We denote the ML estimates as $\hat{\mathbf{u}}=[\hat{f_d},\hat{\theta},\hat{\tau}]^T$.
\par
In practice, $r(t)$ is sampled $N$ times per symbol. This results in a discrete-time version of the LLF as
\begin{equation}
\Lambda(\mathbf{r};\tilde{\nu},\tilde{\theta},\tilde{\varepsilon}) \approx \mathrm{Re}\left[
\sum_{n=0}^{N L_0 -1} e^{-j(2\pi n\tilde{\nu}+\tilde{\theta})} r[n] s_{\tilde{\varepsilon}}^*[n]
\right]
\label{eq:ML-dis}
\end{equation}
where $\nu = f_d T_s/N$, i.e. the normalized frequency offset with respect to the sampling frequency. $r[n]$ and $s_{\varepsilon}[n]$ are the sampled versions of $r(t)$ and $s(t-\varepsilon T_s)$ at $t=n T_s/N$ respectively. Note that $\tilde{\varepsilon}$ is assumed to be zero in the integral limits of  (\ref{eq:ML}) in order to derive (\ref{eq:ML-dis}). This is the main contributor to the approximation in the above given that the sampling frequency is large enough to avoid aliasing.
\begin{figure}[t]
\centering
\includegraphics[width=0.8\linewidth]{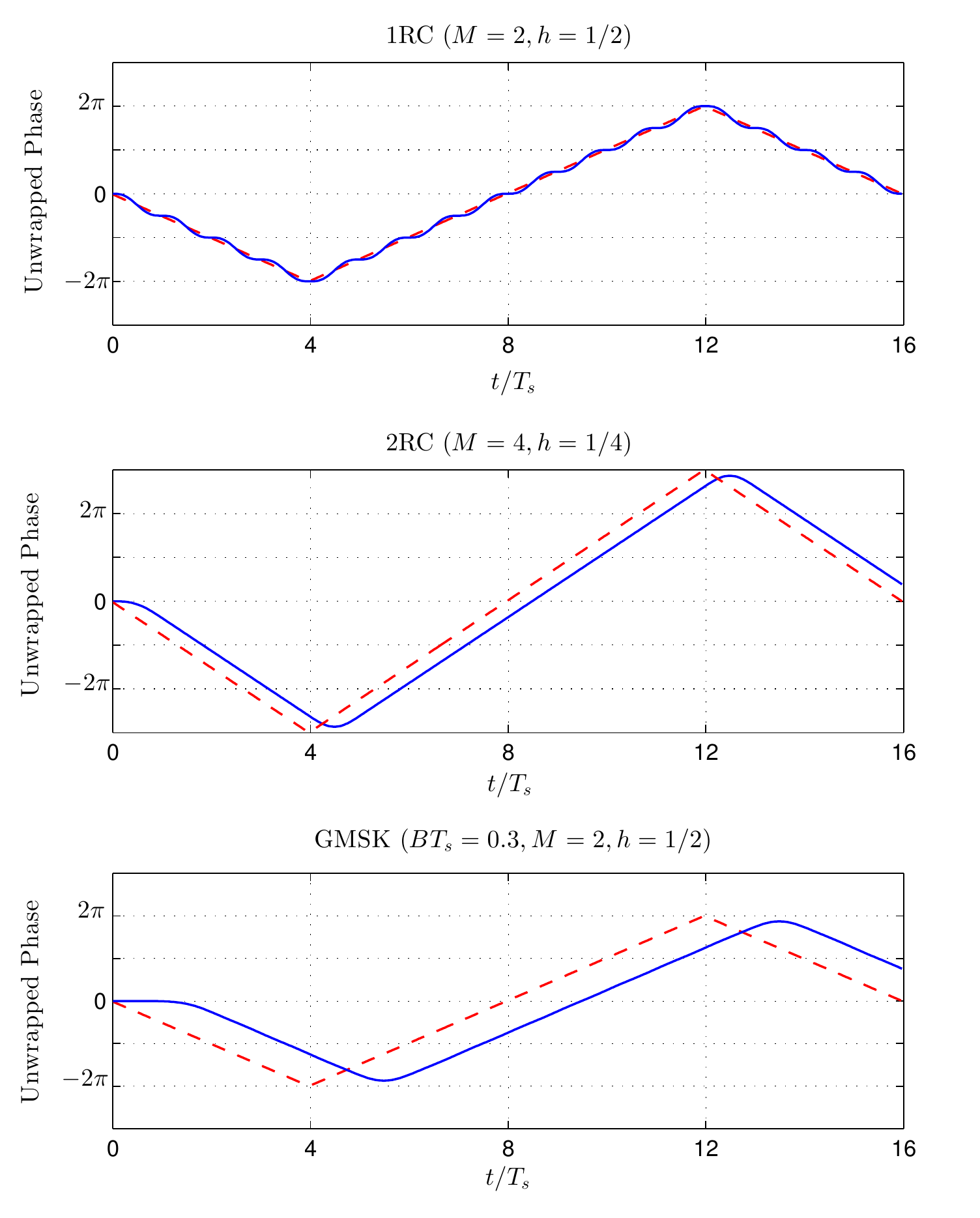}
\caption{The phase response of different CPMs to the optimum training sequence (shown in solid lines). The dashed lines show the response of the same sequence to the 1REC CPM with the same $h$.}
\label{fig:SeqResponse}
\end{figure}
\par 
Based on (\ref{eq:ML-dis}), the maximization of the LLF requires at least a two-dimensional grid search on $(\tilde{\nu},\tilde{\varepsilon})$ in general because both of these parameters are embedded inside the above summation. Therefore, we are interested in a method that decouples $\varepsilon$ and $\nu$. We note that the preamble of Fig. \ref{fig:OptTS}, regardless of its underlying CPM, can be divided into three parts, each of which having the same data symbols. This distinct pattern causes the CPM phase to change with a uniform rate of approximately $\pi h (M-1)$ radians per symbol in the same direction for each part. We have illustrated this fact in Fig. \ref{fig:SeqResponse} by plotting the unwrapped phase response of three different CPMs when preamble of Fig. \ref{fig:OptTS} with $L_0=16$ is utilized. The first signal phase corresponds to the 1RC frequency pulse with binary data symbols and $h=1/2$. Additionally, the partial-response 4-ary 2RC CPM is provided in which $h=1/4$. The GMSK scheme with $BT_s=0.3$ is also included, which is binary, $L=4$ and $h=1/2$. We have compared each case with the phase response of 1REC frequency pulse to the same $\boldsymbol{\alpha}$ and $h$. It is observed that despite the fundamental differences between their frequency pulses, the overall phase response of all CPM signals are approximately similar. More detailed observations can be made as the following: 
\begin{enumerate}
\item 
GMSK and 2RC phase responses follow a straight line within each part similar to the 1REC pulse shape in spite of their bell-shaped pulses. This is due to the overlap of the frequency pulses when the subsequent data symbols are the same, which leads to uniform phase variations. 
\item
The overall phase response is delayed when partial-response CPMs such as 2RC and GMSK are employed. We denote this lag time by $T_l$ which is equivalent to $N_l$ samples. 
\item 
1RC CPM shows the largest deviations from the 1REC phase response because its frequency pulse is full response (non-overlapping) and has the highest peak.  
\end{enumerate}
\par    
Based on the above discussion, we approximate the phase response of any given CPM signal to the optimum preamble $\boldsymbol{\alpha}^*$ with a delayed version of 1REC CPM to $\boldsymbol{\alpha}^*$ and the same $h$ . In fact, the optimum preamble enables us to accurately apply a piecewise linear approximation to the phase of CPM. Therefore, the approximated phase response can be mathematically expressed as
\begin{equation}
\phi(t,\boldsymbol{\alpha}^*)\! \approx\!
\begin{cases}
-(M-1)\pi h\frac{t-T_l}{T_s} & T_l< t \leq \frac{T_0}{4} + T_l\\
(M-1)\pi h\frac{t-T_l-T_0/2}{T_s} & \frac{T_0}{4} + T_l< t  \leq \frac{3T_0}{4} + T_l\\
-(M-1)\pi h\frac{t-T_l-T_0}{T_s} & \frac{3T_0}{4} + T_l< t  \leq T_0 + T_l\\
0 & \mathrm{otherwise}
\end{cases}
\label{eq:phi-approx}
\end{equation}         
where $T_l$ is fixed for a given CPM and is known to the receiver. In the Appendix, it is shown that $T_l = \frac{(L-1)}{2}T_s$ for symmetric $g(t)$, which is the case for rectangular, raised-cosine and Gaussian pulse shapes. In the rest of our discussion, we assume the channel observation starts from $t=T_l$, and hence, we ignore $T_l$. In practice, we can append $\lceil T_l/T_s \rceil$ ``$-(M-1)$ symbols'' to the end of the preamble for partial-response CPMs in order to avoid unwanted variations at the end of the observation interval, which is now shifted by $T_l$. Thus, we use (\ref{eq:phi-approx}) to express $s_\varepsilon[n]$ during the preamble transmission as
\begin{equation}
s_\varepsilon[n]\! \approx\!
\begin{cases}
\!\exp[-j(M\!-\!1)\pi h (\frac{n}{N}\!-\!\varepsilon)]  & \!0< n \leq \frac{NL_0}{4} \\
\!\exp[+j(M\!-\!1)\pi h (\frac{n}{N}\!-\!\frac{L_0}{2}\!-\!\varepsilon)]  & \!\frac{NL_0}{4}< n \leq \frac{3NL_0}{4} \\
\!\exp[-j(M\!-\!1)\pi h (\frac{n}{N}\!-\!L_0\!-\!\varepsilon)]  & \!\frac{3NL_0}{4} < n \leq NL_0. 
\end{cases}
\label{eq:s-approx}
\end{equation}
\par 
We take advantage of the above approximation in order to simplify the LLF and its maximization algorithm. Using (\ref{eq:s-approx}) in (\ref{eq:ML-dis}) results in 
\begin{align}
\Lambda^*(\mathbf{r};\tilde{\nu},\tilde{\theta},\tilde{\varepsilon})\!\approx & \mathrm{Re} \biggl\{\!\!e^{-j\tilde{\theta}}\Bigl[\nonumber
\!\sum_{n=0}^{NL_0/4-1} \!\!\!\!e^{-j2\pi \tilde{\nu} n} r[n] e^{j(M-1)\pi h (n/N-\tilde{\varepsilon})} \Bigr.\biggr. \nonumber\\ 
+& \!\!\sum_{n=NL_0/4}^{3NL_0/4-1} \!\!e^{-j2\pi \tilde{\nu} n} r[n] e^{-j(M-1)\pi h (n/N-L_0/2-\tilde{\varepsilon})} \nonumber\\ 
+&\biggl. \Bigl. \!\!\sum_{n=3NL_0/4}^{NL_0-1} \!\!e^{-j2\pi \tilde{\nu} n} r[n] e^{j(M-1)\pi h (n/N-L_0-\tilde{\varepsilon})}
\Bigr] \biggr\}
\label{eq:LLF-opt}
\end{align}
where $\Lambda^*(\cdot)$ represents the joint LLF given $\boldsymbol{\alpha}^*$. It is evident from (\ref{eq:LLF-opt}) that the symbol timing is now decoupled from the frequency offset and can be moved outside the summations of the LLF. Hence, (\ref{eq:LLF-opt}) can be simplified as
\begin{equation}
\begin{split}
\Lambda^*(\mathbf{r};\tilde{\nu},\tilde{\theta},\tilde{\varepsilon}) &\approx  \\ \mathrm{Re}
\biggl\{
e^{-j\tilde{\theta}}\Bigl[&
e^{-j(M-1)\pi h \tilde{\varepsilon}}\lambda_1(\tilde{\nu})+ 
e^{j(M-1)\pi h \tilde{\varepsilon}}\lambda_2(\tilde{\nu})
\Bigr] \biggr\}
\end{split}
\label{eq:LLF-final}
\end{equation}
where
\begin{equation}
\begin{split}
\lambda_1(\tilde{\nu}) =  &\sum_{n=0}^{NL_0/4-1} e^{-j2\pi \tilde{\nu} n} r[n] e^{j(M-1)\pi h n/N}\\
&+e^{-j(M-1)\pi h L_0}\sum_{n=3NL_0/4}^{NL_0-1}\!\!\! e^{-j2\pi \tilde{\nu} n} r[n] e^{j(M-1)\pi h n/N}
\end{split}
\label{eq:lambda1}
\end{equation}
and
\begin{equation}
\lambda_2(\tilde{\nu})\!=
e^{j(M-1)\pi h L_0/2}\!\sum_{n=NL_0/4}^{3N L_0/4-1} \!e^{-j2\pi \tilde{\nu} n} r[n] e^{-j(M-1)\pi h n/N}.
\label{eq:lambda2}
\end{equation}
\par 
As the estimation parameters are now decoupled, the maximization of the LLF becomes straightforward. Let us proceed by denoting the term in (\ref{eq:LLF-final}) which corresponds to symbol timing and frequency offset as
\begin{equation}
\Gamma(\tilde{\nu},\tilde{\varepsilon}) =
e^{-j(M-1)\pi h \tilde{\varepsilon}}\lambda_1(\tilde{\nu})+
e^{j(M-1)\pi h \tilde{\varepsilon}}\lambda_2(\tilde{\nu}).
\label{eq:gamma}
\end{equation}
It is easily seen that for any value of $(\tilde{\nu},\tilde{\varepsilon})$, $\Lambda^*(\cdot)$ is maximized by choosing $\tilde{\theta}$ such that it rotates $\Gamma(\tilde{\nu},\tilde{\varepsilon})$ towards the real axis, i.e.,
\begin{equation}
\tilde{\theta} = \arg\{\Gamma(\tilde{\nu},\tilde{\varepsilon})\}.
\label{eq:phi-tilde}
\end{equation}
which reduces the LLF to $|\Gamma(\tilde{\nu},\tilde{\varepsilon})|$. Thus, the ML estimates of $\tilde{\nu}$ and $\tilde{\varepsilon}$ are found by maximizing
\begin{equation}
\begin{split}
|\Gamma(\tilde{\nu},\tilde{\varepsilon})|^2 \!= \!|\lambda_1(\tilde{\nu})|^2\!\!+\!\!|\lambda_2(\tilde{\nu})|^2
\!\!+\!\!2\mathrm{Re}[ e^{-j2(M-1)\pi h \tilde{\varepsilon}}\!\lambda_1(\tilde{\nu})\lambda_2^*(\tilde{\nu})]
\end{split}
\label{eq:gamma2}
\end{equation}
with respect to $(\tilde{\nu},\tilde{\varepsilon})$. The first two terms on the right-hand side of (\ref{eq:gamma2}) do not depend on $\tilde{\varepsilon}$. Using a similar argument as $\tilde{\theta}$, the third term is maximized by selecting $\tilde{\varepsilon}$ according to
\begin{equation}
\tilde{\varepsilon} = \frac{\arg\{\lambda_1(\tilde{\nu})\lambda_2^*(\tilde{\nu})\}}{2(M-1)\pi h}
\label{eq:eps-tilde}
\end{equation}
so that the term inside the real part operator in (\ref{eq:gamma2}) becomes purely real and equal to $|\lambda_1(\tilde{\nu})\lambda_2^*(\tilde{\nu})|$. Therefore, the maximization of the LLF is now a one-dimensional problem that results in the ML estimate of $\nu$. This can be expressed mathematically in the form of
\begin{equation}
\hat{\nu} = \underset{\tilde{\nu}}{\arg\!\max}\left\{ X(\tilde{\nu}) = |\lambda_1(\tilde{\nu})|+|\lambda_2(\tilde{\nu})|\right\}
\label{eq:fd-hat}
\end{equation} 
which in turn leads to the ML estimates of the normalized symbol timing and phase offset via
\begin{equation}
\hat{\varepsilon} = \frac{\arg\{\lambda_1(\hat{\nu})\lambda_2^*(\hat{\nu})\}}{2(M-1)\pi h}
\label{eq:eps-hat}
\end{equation}
and
\begin{equation}
\hat{\theta} = \arg\left\{e^{-j(M-1)\pi h \hat{\varepsilon}}\lambda_1(\hat{\nu})+
e^{j(M-1)\pi h \hat{\varepsilon}}\lambda_2(\hat{\nu})\right\}.
\label{eq:theta-hat}
\end{equation}
respectively.
\subsection{Implementation of the Frequency Offset Estimator}
In the previous section, we derived simple closed-form expressions for estimation of phase and symbol timing. However, the frequency offset estimation requires computing the maximum of a one-dimensional function as defined in (\ref{eq:fd-hat}). $\lambda_1(\nu)$ and $\lambda_2(\nu)$ have the form of Fourier transforms of $r(t)$ and should be expected to have fluctuations due to the presence of noise, which results in several local maxima. Thus, a grid search is inevitable in order to find the correct frequency offset with confidence. 
\par
According to (\ref{eq:lambda1}) and (\ref{eq:lambda2}), computations of $\lambda_1(\nu)$ and $\lambda_2(\nu)$ require a different number of summations with different limits. In order to make both of them consistent, we define two new signals, i.e. $r_1[n]$ and $r_2[n]$, such that
\begin{equation}
r_1[n]=
\begin{cases}
r[n] & 0\leq n <N L_0/4 \\
\exp[-j(M-1)\pi h L_0]r[n] & 3NL_0/4\leq n <N L_0\\
0 & \text{otherwise}
\end{cases}
\label{eq:r1}
\end{equation}
and
\begin{equation}
r_2[n]\!=\!
\begin{cases}
\exp[j(M-1)\pi h L_0/2]r[n] & \!NL_0/4 \leq \!n\!< 3N L_0/4\\
0 & \!\text{otherwise.}
\end{cases}
\label{eq:r2}
\end{equation} 
The above modifications to $r[n]$ leads to similar forms for $\lambda_1(\nu)$ and $\lambda_2(\nu)$, where each one requires computation of one summation with $NL_0$ terms, i.e.,
\begin{equation}
\lambda_1(\tilde{\nu}) = \sum_{n=0}^{NL_0-1} r_1[n]e^{j(M-1)\pi h n/N}
e^{-j2\pi n \tilde{\nu}}
\label{eq:lambda1-dis}
\end{equation}
and
\begin{equation}
\lambda_2(\tilde{\nu}) = \sum_{n=0}^{NL_0-1} r_2[n]e^{-j(M-1)\pi h n/N}
e^{-j2\pi n \tilde{\nu}.}
\label{eq:lambda2-dis}
\end{equation}
\par 
The computations of (\ref{eq:lambda1-dis}) and (\ref{eq:lambda2-dis}) for different $\tilde{\nu}$ values resemble the discrete Fourier transform (DFT) operation, where $\tilde{\nu}$ is replaced by trial discrete frequencies. These operations can be performed efficiently using the fast Fourier transform (FFT). The FFT size will be equal to the summation length assuming $NL_0$ is a power of two. This process results in trial values for $\lambda_1(\tilde{\nu})$ and $\lambda_2(\tilde{\nu})$ such that $\tilde{\nu} \in [0,1/NL_0,\dots,(NL_0-1)/NL_0]$, which are then inserted in (\ref{eq:fd-hat}) in order to find $\hat{\nu}$. Therefore, the frequency offset estimate requires two FFTs of the same size.
\par 
The frequency estimation performance is limited by the resolution of the FFT operations, i.e. the distance between the discrete frequency components. A low resolution estimate may cause a ripple effect on the estimation performance of other parameters. In order to increase the accuracy of the frequency estimate, two approaches are considered. The first approach is to zero pad the FFT operands in (\ref{eq:lambda1-dis}) and (\ref{eq:lambda2-dis}) such that both FFTs have a size of $N_f = K_fNL_0$ where $K_f$ is a power of two. This procedure results in a frequency resolution of $1/K_fL_0$ with respect to the symbol rate. The second approach is to employ an interpolator in order to estimate the true maximum of (\ref{eq:fd-hat}) between the discrete frequency values. In \cite{Gasior2004}, it was shown that the Gaussian interpolator is superior to a parabolic one in terms of improving FFT resolution. The only added complexity is an extra look-up table for computation of the logarithm function. The Gaussian interpolation can be expressed as
\begin{equation}
\hat{\nu} = \hat{\nu}_0 + \frac{1}{2K_fNL_0}\frac{\log X(\hat{\nu}_{-1})- \log X(\hat{\nu}_{1}) }{\log X(\hat{\nu}_{-1}) + \log X(\hat{\nu}_{1}) - 2\log X(\hat{\nu}_{0})}
\label{eq:interpolator}
\end{equation}
where $\hat{\nu}_0$ represents the maximizing frequency resulting from (\ref{eq:fd-hat}). $\hat{\nu}_{-1}$ and $\hat{\nu}_1$ denote the discrete frequency components immediately before and after $\hat{\nu}_0$ respectively in terms of the FFT operation. The above operation can be regarded as a \textit{fine search} while FFTs perform a \textit{coarse search} on the frequency offset. 
\par
Based on DFT properties, FFT operations are periodic with a period of $NL_0$. Therefore, values of $1/2 \leq \hat{\nu} < 1$ represent negative frequency offsets, and hence, $\hat{\nu}$ is estimated over $[-1/2,1/2)$. This limits the frequency estimation range to
\begin{equation}
-\frac{N}{2T_s}\leq \hat{f}_d < \frac{N}{2T_s}
\end{equation}
\begin{figure}[t]
\includegraphics[width=\linewidth]{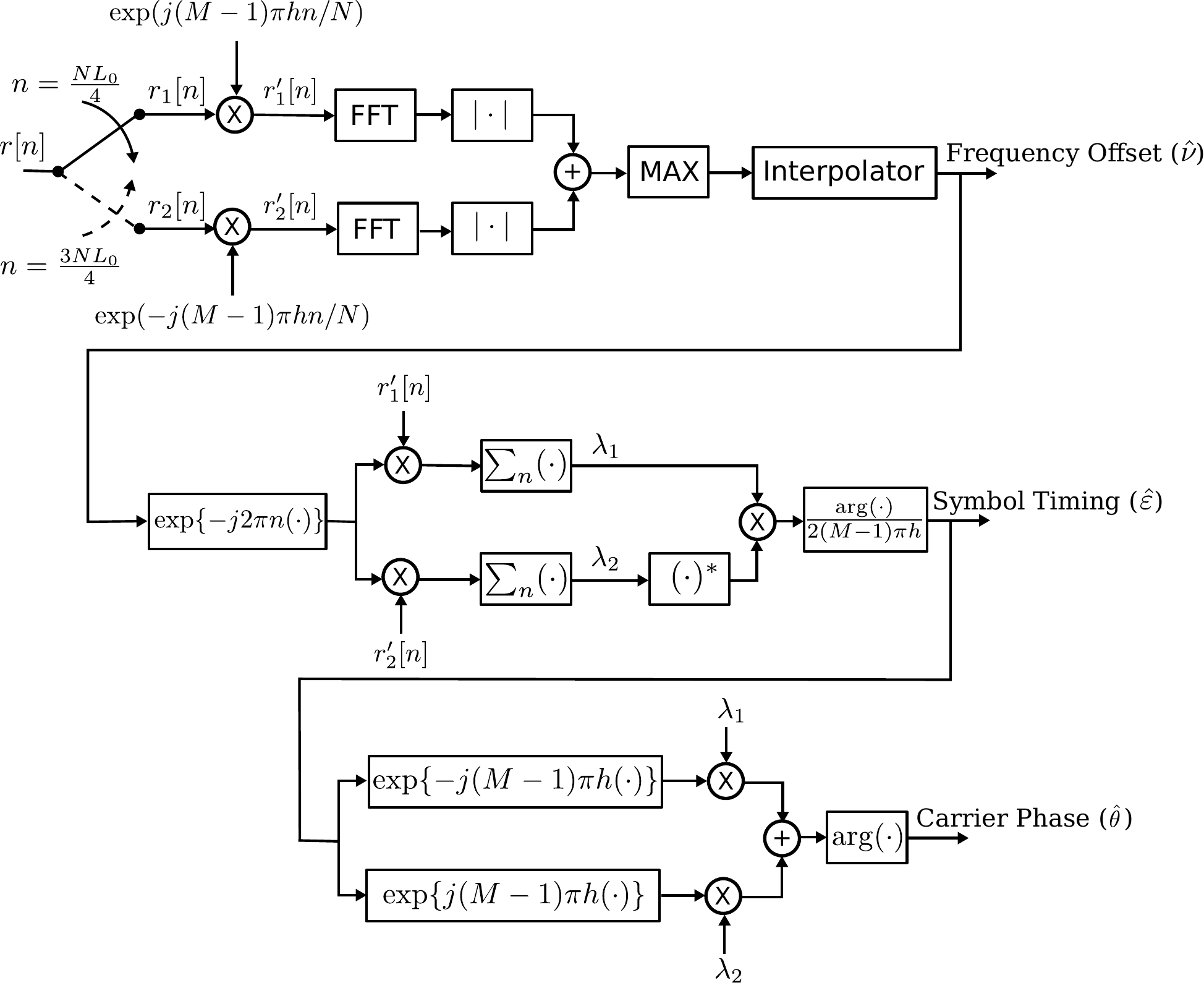}
\caption{Block diagram of the feedforward joint frequency offset, symbol timing and carrier phase estimator.}
\label{fig:estimator}
\end{figure}
which can be increased by increasing the sampling frequency. Therefore, the proposed algorithm can easily handle applications in which the frequency offset is greater than the symbol rate.
\par
The final design for our feedforward joint frequency offset, symbol timing and carrier phase estimator is illustrated in Fig. \ref{fig:estimator}. Based on (\ref{eq:r1}) and (\ref{eq:r2}), $r_1[n]$ and $r_2[n]$ should be multiplied by $\exp[-j(M-1)\pi h L_0]$ and $\exp[j(M-1)\pi h L_0/2]$ respectively. However, we have not shown this in Fig. \ref{fig:estimator} for the sake of clarity, and because the aforementioned factors are basically equal to one in our examples. Based on the above block diagram, the joint estimator requires $2NL_0+3$ complex multiplications, $NL_0$ real multiplications, $NL_0+1$ complex additions and $K_fNL_0$ real additions. These exclude blocks such as FFT, interpolation, $|\cdot|$, $\exp(\cdot)$, and $\arg(\cdot)$ as their complexity depends on their implementation method. 
\section{Frame Synchronization}
So far, we have assumed the carrier and timing synchronization algorithm have the knowledge of the SoS within $\pm T_s/2$, which has to be carried out by the frame synchronization algorithm. In this work, we decompose the frame synchronization into two tasks: \textit{SoS detection} and \textit{SoS Estimation}. The SoS detector determines the arrival of a new burst such that the preamble is located within an \textit{observation} or uncertainty window. The SoS estimation algorithm then tries to find the exact location of the SoS within that window. Using a reverse approach, we initially derive the SoS estimation algorithm. Based on its results, we propose a simple SoS detection algorithm. It should be mentioned that if the observation window does not contain the whole preamble due to SoS detection failures, the SoS estimation results are no longer reliable, and hence, an entire burst might be missed. 
\subsection{SoS Estimation Algorithm}
\begin{figure}[t]
\centering
\includegraphics[width=0.7\linewidth]{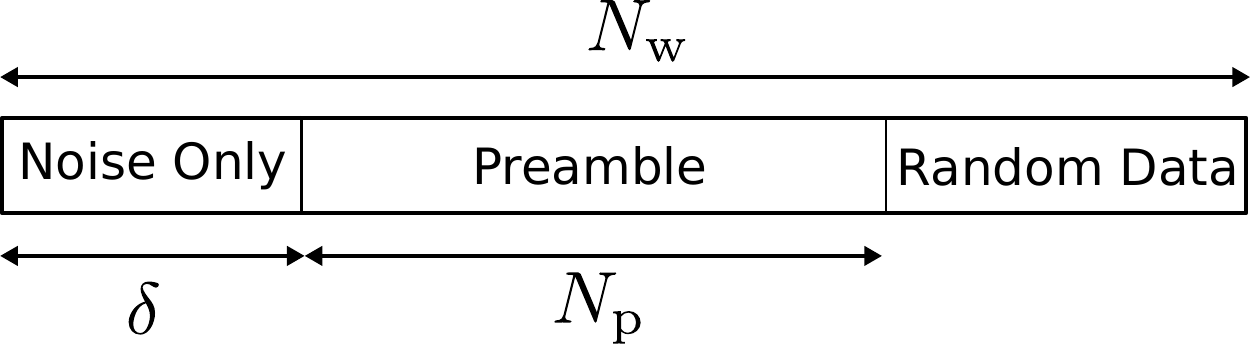}
\caption{The observation window for the SoS estimation algorithm.}
\label{fig:SosEstimator}
\end{figure}

The framework for the SoS estimation algorithm is depicted in Fig. \ref{fig:SosEstimator} where an observation window of $N_w$ samples is considered. The first $\delta$ samples contain only WGN which correspond to the guard interval prior to the beginning of signal transmission. Note that $\delta$ differs from $\mu$ in (\ref{eq:delay}) where the latter one represents the overall delay from the transmitter to the receiver. It is followed by $N_p$ samples of the preamble. Finally, there are $N_w-\delta-N_p$ samples which are assumed to be generated from a random CPM signal, and are associated with the UW and/or payload portion of the burst. The SoS estimation algorithm attempts to find the best estimate of $\delta$ according to the above observation window. 
\par
The received and sampled signal within the observation window can be expressed as
\begin{equation}
r[n]=
\begin{cases}
w[n] & 0 \leq n < \delta\\
e^{j(2\pi\nu n + \theta)}s[n-\delta] + w[n] & \delta \leq n < N_w
\end{cases}
\label{eq:rx-window}
\end{equation}
where $w[n]$ is complex white Gaussian random sequence with a variance of $\sigma^2=N(E_s/N_0)^{-1}$. Additionally, we have assumed $T_s=1$ and $|s[n]|=1$. It should be noted that $\theta$ in (\ref{eq:rx-window}) is different from its value in (\ref{eq:cpm-received}) due to the frequency offset and different reference points. Finally, we denote the values of $r[n]$ within the observation window by $\mathbf{r}$.
\par 
Based on the ML rules, the best estimate of $\delta$ is the value which maximizes the likelihood function $p(\mathbf{r};\delta)$. However, let us first consider the likelihood function as a function of all unknown parameters, i.e., 
\begin{equation}
\begin{split}
p(\mathbf{r};\delta,\nu,\theta,\boldsymbol{\alpha}_d)\! =& \frac{1}{(\pi\sigma^2)^{N_w}}
\exp\left(-\frac{1}{\sigma^2}\sum_{n=0}^{\delta-1} |r[n]|^2 \right)\\
&. \!\exp\!\!\left(\!-\frac{1}{\sigma^2}
\!\!\sum_{n=\delta}^{N_w-1}\!\! |r[n]\!-\!s[n-\delta]e^{j(2\pi\nu n + \theta)}|^2 
\!\right)
\end{split}
\label{eq:lf-dis-orig}
\end{equation}
where $\boldsymbol{\alpha}_d$ represents the random data sequence (uniformly distributed among all sequences of that length) in the non-preamble portion of $s[n]$ and $\mathbf{r}$ is the received signal vector. If we omit constant factors in the likelihood function, it becomes
\begin{equation}
\begin{split}
p(\mathbf{r};\delta,\nu,\theta,\boldsymbol{\alpha}_d) \!=& 
\exp\left( \frac{\delta-N_w}{\sigma^2}\right)\\ 
&.\!\exp\!\!\left(\!\frac{2}{\sigma^2}\!\!\sum_{n=\delta}^{N_w-1} \!\!\mathrm{Re}\! \left\{\!r^*[n]s[n\!-\!\delta]e^{j(2\pi\nu n + \theta)}\right\}\!\!\right)\!\!.
\end{split}
\label{eq:lf-dis-full}
\end{equation}
In order to compute $p(\mathbf{r};\delta)$ from (\ref{eq:lf-dis-full}), we must either estimate or average out the \textit{nuisance} parameters, i.e. $\nu$, $\theta$ and $\boldsymbol{\alpha}_d$, which is not trivial due to the form of the above function. Instead, we initially approximate the exponential function with its second degree Taylor's series in the neighborhood of zero, i.e.,
\begin{equation}
\begin{split}
 p&(\mathbf{r};\delta,\nu,\theta,\boldsymbol{\alpha}_d)\!\!\!\!\!\!\!\!\!\!\!\\
 &\approx C(\delta)\!
 \Biggl(\!1\! + \frac{2}{\sigma^2}\sum_{n=\delta}^{N_w-1} \mathrm{Re} \left\{r^*[n]s[n-\delta]e^{j(2\pi\nu n + \theta)}\right\}  \\+&
\frac{1}{\sigma^4}\!\!\!\sum_{n=\delta}^{N_w-1}\! \sum_{m=\delta}^{N_w-1} 
\!\!\mathrm{Re}\! \left\{\!r^*[n]r^*[m]s[n\!-\!\delta]s[m\!-\!\delta]e^{j(2\pi\nu (m+n) + 2\theta)}\!\right\}\\+&
\frac{1}{\sigma^4}\!\!\!\sum_{n=\delta}^{N_w-1}\! \sum_{m=\delta}^{N_w-1} 
\!\!\mathrm{Re}\! \left\{\!r^*[n]r[m]s[n\!-\!\delta]s^*[m\!-\!\delta]e^{j(2\pi\nu (n-m))}
\!\right\}\Biggr)
\end{split}
\label{eq:lf-dig2}
 \end{equation}
 where $C(\delta)$ represents $\exp( \frac{\delta-N_w}{\sigma^2})$ in (\ref{eq:lf-dis-full}), which is not a function of the nuisance parameters. However, we avoid using it in its original form because it can be very small and adversely affect the approximated likelihood function. Nevertheless, we will propose an approximation for $C(\delta)$ once the final form of the likelihood function becomes available.
 \par 
 Assuming $\theta$ is uniformly distributed over $[-\pi,\pi)$, it can be eliminated from the likelihood function by averaging (\ref{eq:lf-dig2}) over $\theta$, i.e.,
\begin{equation}
\begin{split}
 &p(\mathbf{r};\delta,\nu,\boldsymbol{\alpha}_d) =
 \frac{1}{2\pi}\int_{-\pi}^{\pi} p(\mathbf{r};\delta,\nu,\theta,\boldsymbol{\alpha}_d) \, d\theta\\
 &\approx  C(\delta) \frac{1}{\sigma^4}\sum_{n=\delta}^{N_w-1} \sum_{m=\delta}^{N_w-1} 
 \mathrm{Re} \biggl \{
r^*[n]r[m]s[n-\delta]\\
& \qquad\qquad\qquad\qquad\qquad\qquad .s^*[m-\delta]e^{j(2\pi\nu (n-m))}\biggr\}
.\end{split}
 \label{eq:lf-tfd}
\end{equation}
Note that we have neglected 1 in (\ref{eq:lf-dig2}) because it is much smaller than the fourth term especially when noise variance is small. We also omit $1/\sigma^4$ from the above as it is a constant factor. If we denote $d=m-n$ in (\ref{eq:lf-tfd}), it can be rearranged as
\begin{equation}
\begin{split}
p&(\mathbf{r};\delta,\nu,\boldsymbol{\alpha}_d) \approx
C(\delta)\! 
\left(\sum_{n=\delta}^{N_w-1} |r[n]|^2 \right.\\
+ 2\!\!\!\!\!\! & \left.\sum_{d=1}^{N_w-\delta-1} \!\!\!\!\!\mathrm{Re}\!\left\{\!\!e^{-j2\pi d\nu}\!\! \!\sum_{n=\delta}^{N_w-d-1}
\!\!\!\!\! r^*[n]r[n\!+\!d]s[n\!-\!\delta]s^*[n\!+\!d\!-\!\delta]
\right\}\!\right)
\end{split}
\label{eq:lf-tfd2}
\end{equation}
which allows us to investigate signal correlation due to the presence of random $\boldsymbol{\alpha}_d$. The computation of $E_{\boldsymbol{\alpha}_d}\{p(\mathbf{r};\delta,\nu,\boldsymbol{\alpha}_d)\}$ leads us to compute $E_{\boldsymbol{\alpha}_d}\{s[n-\delta]s^*[n+d-\delta]\}$ which, in our problem, is 
\begin{equation}
\begin{split}
& E_{\boldsymbol{\alpha}_d}\{s[n-\delta]s^*[n+d-\delta]\}\\ &=
\begin{cases}
s[n-\delta]s^*[n+d-\delta] & \delta \leq n < N_p+\delta-d \\
0 & N_p+\delta-d \leq n <N_p + \delta \\
R_{ss}(d) & n \geq N_p + \delta
\end{cases}
\end{split}
\label{eq:ExpData}
\end{equation}
where $R_{ss}(d)$ is the autocorrelation function of the CPM signal normalized to the sample duration. $R_{ss}(d)$ can be computed numerically as described in \cite[p. 208]{Proakis}. The first case in (\ref{eq:ExpData}) corresponds to the preamble, which has no randomness. The second case is zero because $s[n-\delta]$ is deterministic whereas $s^*[n+d-\delta]$ is generated by the random data and its expected value is zero. Therefore, taking the expected value of (\ref{eq:lf-tfd2}) with respect to $\boldsymbol{\alpha}_d$ results in
\begin{equation}
\begin{split}
p(\mathbf{r};\delta,\nu)& \approx
C(\delta) \Bigg(
\sum_{n=\delta}^{N_w-1} |r[n]|^2 +
2\!\sum_{d=1}^{N_p-1} \!\!\mathrm{Re}\Bigg\{e^{-j2\pi d\nu}\\& 
.\Bigg(\sum_{n=\delta}^{N_p+\delta-d-1} r^*[n]r[n+d]s[n-\delta]s^*[n+d-\delta] \\ &\quad\quad+ 
 \!R_{ss}(d)\sum_{n=N_p+\delta}^{N_w-d-1} r^*[n]r[n+d]
\Bigg)\Bigg\}
\Bigg).
\end{split}
\label{eq:lf-tf}
\end{equation}
In general, CPM autocorrelation function becomes zero for lag times greater than $LT_s$. Therefore, $R_{ss}(d)$ is zero except for its first few values.
\par 
The last step to obtain $p(\mathbf{r};\delta)$ is removing $\nu$ from (\ref{eq:lf-tf}). It can be verified that averaging (\ref{eq:lf-tf}) with respect to $\nu$, uniformly distributed over [-0.5,0.5), completely eliminates the second summation. This indeed results in a poor ML estimate for $\delta$ because it ignores the knowledge of the known preamble. A better approach is to estimate $\nu$ by maximizing the second summation in the above. However, a closed-form solution seems to be unavailable due to the range of $d$. Instead, we can derive different estimates $\hat{\nu}_d$ based on single terms inside the summation via
\begin{equation}
\begin{split}
\hat{\nu}_d = \frac{1}{2\pi d}\arg\Biggl\{
&\sum_{n=\delta}^{N_p+\delta-d-1} \!\!\! r^*[n]r[n+d]s[n-\delta]s^*[n+d-\delta]\\&+
R_{ss}(d)\!\!\sum_{n=N_p+\delta}^{N_w-d-1}\!\!\! r^*[n]r[n+d]
\Biggr\}.
\end{split}
\label{eq:freq-est}
\end{equation}
The above method is the basis for some well-known carrier frequency estimation algorithms, such as \cite{Fitz}. If we use $\hat{\nu}_d$ values and plug them back into (\ref{eq:lf-tf}), the likelihood function becomes independent of the frequency offset. Thus,
 \begin{equation}
 \begin{split}
 p&(\mathbf{r};\delta) \approx
 C(\delta) \Bigg(
 \sum_{n=\delta}^{N_w-1} |r[n]|^2  \\&+ 
 2\!\sum_{d=1}^{N_p-1} \Bigg| 
 \sum_{n=\delta}^{N_p+\delta-d-1} r^*[n]r[n+d]s[n-\delta]s^*[n+d-\delta]\\
 &\qquad\qquad\quad+R_{ss}(d)\sum_{n=N_p+\delta}^{N_w-d-1} r^*[n]r[n+d]
 \Bigg|
 \Bigg)
 \end{split}
 \label{eq:lf-t}
 \end{equation}
 which must be maximized with respect to $\delta$ in order to derive $\hat{\delta}$. 
 \par 
 The computational complexity of (\ref{eq:lf-t}) can be reduced by truncating the summation over $d$. This results in a sub-optimum, reduced-complexity estimator, i.e.,
 \begin{equation}
 \begin{split}
 &\hat{\delta} = \underset{\tilde{\delta}}{\arg\!\max} \,\, \Bigg\{ C(\tilde{\delta})\\ 
 &.\Bigg(
 \!\! \sum_{n=\tilde{\delta}}^{N_w-1} \!\!|r[n]|^2 +
  2\!\!\sum_{d=1}^{D}\! \Bigg|\!\! 
  \sum_{n=\tilde{\delta}}^{N_p+\tilde{\delta}-d-1}\!\!\!\!\! r^*[n]r[n\!+\!d]s[n\!-\!\tilde{\delta}]s^*[n\!+\!d\!-\!\tilde{\delta}]\\
  &\qquad\qquad\qquad\qquad\quad+R_{ss}(d)\sum_{n=N_p+\tilde{\delta}}^{N_w-d-1} r^*[n]r[n+d]
  \Bigg|
  \Bigg)
  \Bigg\}
  \end{split}
  \label{eq:delta-est}
 \end{equation}
 where $1 \leq D <N_p$ is a design parameter, which allows a trade-off between complexity and performance. In fact, if we assume $R_{ss}(d) = 0$ for $d \neq 0$, it is observed that computation of the argument of (\ref{eq:delta-est}) requires $D(2N_p-D-1)+N_w-N_p$ complex multiplications, 2 real multiplications, $D(N_p-\frac{D+1}{2})+N_w-N_p$ complex additions, $D$ real additions, and $D$ computations of the absolute value. Therefore, the complexity of proposed algorithm is approximately a linear function of $D$.
 \par  
 As mentioned earlier, $C(\delta)$ needs to be adjusted based on the final form of the likelihood function. We note that (\ref{eq:lf-t}) is dominated by the summation over $d$. If we ignore $R_{ss}(d)$ due to its short length, the computations inside the absolute value are performed over a sliding window that covers the hypothetical preamble. If this window is shifted to the left by one sample, one signal-plus-noise sample will be replace by one noise-only sample, which has a smaller energy compared to the former one. However, shifting the window to the right replaces it with a different signal-plus-noise sample. Therefore, we expect $p(\mathbf{r};\delta+1) > p(\mathbf{r};\delta-1)$, if $\delta$ is its true value. This makes the likelihood test biased, i.e. $\hat{\delta}$ is more likely to tend towards $\delta+1$ than $\delta-1$. We introduce a simple solution to this issue by proposing
 \begin{equation}
 C(\delta) \triangleq (N_w-\delta)^q
 \label{eq:cdelta}
 \end{equation}
  where $q \geq 0$ is another design parameter which has to be chosen according to $D$. As we will see in the simulation results, $q=1$ is a good choice for the full-complexity estimator, i.e. $D=N_p-1$, while it has to be reduced for smaller values of $D$.
  \par
  Choi and Lee \cite{Lee2002a} have presented a ML frame synchronization algorithm through a different path for PSK signals where the preamble is surrounded by random data. Despite similarities to (\ref{eq:delta-est}), our estimator addresses a different scenario in which the preamble is preceded by the noise-only samples so that $C(\delta)$ was introduced. Additionally, the memory of CPM signals is handled via the presence of $R_{ss}(d)$. Finally, it should be mentioned that each of the  summations $\sum_n r^*[n]r[n+d]s[n-\tilde{\delta}]s^*[n+d-\tilde{\delta}]$ in (\ref{eq:delta-est}) is referred to as a \textit{double-correlation} in \cite{Lee2002a}.
\subsection{SoS Detection Algorithm}
As the last piece of our synchronization algorithm, we present a simple ML detection algorithm, which is closely related to our previous discussion. Let us assume a receiver which collects vectors of $N_p$ samples using a sliding window. We denote this vector by $\mathbf{r}_p$. Additionally, consider two hypotheses $\mathcal{H}_0$ and $\mathcal{H}_1$. $\mathcal{H}_0$ is the hypothesis where the entire vector of samples in $\mathbf{r}_p$ are noise-only samples, which happens when no burst is received. On the other hand, $\mathcal{H}_1$ is the hypothesis where $\mathbf{r}_p$ is perfectly aligned with the preamble. We can distinguish these two hypotheses by performing a likelihood ratio test (LRT) according to
\begin{equation}
L(\mathbf{r}_p) = \frac{p(\mathbf{r}_p;\mathcal{H}_1)}{p(\mathbf{r}_p,\mathcal{H}_0)}
\overset{\mathcal{H}_1}{\underset{\mathcal{H}_0}{\gtrless}} \gamma
\label{eq:LrtDef}
\end{equation}
where $p(\mathbf{r}_p;\mathcal{H}_i)$ is the likelihood function under $\mathcal{H}_i$. Based on the above test, $\mathcal{H}_1$ is selected when $L(\mathbf{r}_p)$ is greater than a threshold $\gamma$. Otherwise, we select $\mathcal{H}_0$, i.e. no preamble is present.
\par
Obviously, several other hypotheses also occur in between these two in which $\mathbf{r}_p$ contains only a fraction of the preamble, i.e. a \textit{mixed-signal} scenario. However, we note that the LRT is performed on a sliding window, which makes it a generalized LRT (GLRT). In this manner, the unknown time delay is being estimated at the same time as the preamble is detected. The point at which the above test exceeds the threshold indicates the presence of the preamble and an estimate of the SoS. Later, the SoS estimator improves this estimate by going over an additional $N_p$ samples to find the peak when it considers the full structure of a burst, i.e. the preamble is followed by CPM signal rather than noise.
\par
The likelihood ratio of (\ref{eq:LrtDef}) can be easily obtained from (\ref{eq:lf-t}). In fact, $p(\mathbf{r}_p;\mathcal{H}_1)$ becomes equal to $p(\mathbf{r};\delta)$ when $\delta=0$ and $N_w=N_p$. We also note that we can multiply $p(\mathbf{r}_p;\delta=0)$ by $\exp(-\frac{1}{\sigma^2}\sum_{n=0}^{N_p-1}|r[n]|^2)$ because it is not a function of $\delta$. The latter factor is basically equal to $p(\mathbf{r}_p;\mathcal{H}_0)$. Thus, the likelihood test can be approximated by
\begin{equation}
\begin{split}
& L(\mathbf{r}_p) = \frac{p(\mathbf{r}_p;\delta=0)p(\mathbf{r}_p;\mathcal{H}_0)}{p(\mathbf{r}_p;\mathcal{H}_0)}\\
&\approx\!\!\!
\sum_{n=0}^{N_p-1} 
\!\!\left|r[n]\right|^2 \!+\! 
2\!\!\!\sum_{d=1}^{Np-1}\!\left|\sum_{n=0}^{Np-d-1} \!\!\!\!r^*[n]r[n+d]s[n]s^*[n+d] 
\right|\gtrless \gamma'.
\end{split}
\label{eq:LrtApprox}
\end{equation}
Similar to (\ref{eq:delta-est}), we propose a reduced-complexity test, i.e.
\begin{equation}
L_{D'}(\mathbf{r}_p) \triangleq
\sum_{d=1}^{D'}\left|\sum_{n=0}^{Np-d-1} r^*[n]r[n+d]s[n]s^*[n+d] 
\right| \gtrless \gamma_{D'}.
\label{eq:LrtTruncated}
\end{equation}
where $1 \leq D' < N_p$ is a design parameter and $\gamma_{D'}$ represents the test threshold for a given $D'$. It can be verified that computation of (\ref{eq:LrtTruncated}) requires $D'(2N_p - D'-1)$ complex multiplications, $D'(N_p - \frac{D'+1}{2})$ complex additions, $D'$ real additions, and $D'$ absolute value functions. Moreover, the majority of computations in SoS detection and SoS estimation are the same, which allows a high degree of resource sharing between these two blocks.
\par 
The threshold $\gamma_{D'}$ can be chosen based on the Neyman-Pearson criterion \cite{Kay1993a} in which the probability of \textit{false alarm} is fixed. Here, the probability of false alarm is defined as $P_{\text{FA}}=\text{Pr}\{L_{D'}(\mathbf{r}_p) > \gamma_{D'}|\mathcal{H}_0 \}$. Once the threshold is chosen, the probability of \textit{missed detection} can be calculated via $P_{\text{MD}} = \text{Pr}\{L_{D'}(\mathbf{r}_p) < \gamma_{D'}|\mathcal{H}_1\}$. The probability of correct detection is $P_{\text{D}}=1-P_{\text{MD}}$. Exact closed-form expressions for $P_{\text{FA}}$ and $P_{\text{D}}$ may not be realized due to the magnitude operators and multiplications in (\ref{eq:LrtTruncated}). For instance, if we denote the output of each double-correlation as a random variable, i.e., $X_d = \sum_{n=0}^{N_p-d-1} r^*[n]r[n+d]s[n]s^*[n+d]$, a simple yet acceptable (for large $N_p-d$) approximation is to consider $X_d$ as a complex Gaussian random variable (RV). This forces $|X_d|$ to become a Rayleigh RV under $\mathcal{H}_0$ and Rician RV under $\mathcal{H}_1$ due to the presence of signal. Thus, $L_{D'}(\mathbf{r}_p)$ can be approximated as sum of Rayleigh or Rician RVs depending on the hypothesis. In \cite{Hu2005,Hu2005a}, approximate cumulative distribution functions (CDFs) are provided for such RVs. However, our investigations show that the approximation error is considerable because we are interested in regions where $P_{\text{FA}}$ and $P_{\text{MD}}$ are very low.
Therefore, we resort to Monte-Carlo simulations with a large sample size in order to compute these probabilities, $\gamma_{D'}$, and the receiver operating characteristic (ROC).
\section{Results and Discussion}
\subsection{Approximation Error}
In this section, we study the error in representing the CPM phase during preamble transmission using (\ref{eq:phi-approx}). For the sake of clarity, we denote the original CPM phase by $\phi(t)$, and its approximated value in (7) by $\phi'(t)$. Therefore, the representation error is $e(t) = \phi(t)-\phi'(t)$. We define the approximation error as the ratio of the energy in the error ($E_e$) to the signal energy during  the preamble transmission, i.e.,
\begin{equation}
e_a \triangleq \frac{\int_{0}^{T_0}|e^{j\phi(t)}-e^{j\phi'(t)}|^2 \,dt}{\int_{0}^{T_0}\! |e^{j\phi(t)}|^2\, dt}=\frac{\int_{0}^{T_0}\!|1-e^{je(t)}|^2\,dt}{L_0 T_s}=\!\frac{E_e}{L_0T_s}.
\label{eq:approx-error}
\end{equation}
In the above, $|1-e^{je(t)}|^2$ can be approximated by $e^2(t)$ for small values of $|e(t)|$, which gives us an approximate value of  $E_e\approx \int_{0}^{T_0} e^2(t) \, dt$. The computation of $E_e$ for full response CPMs is 
\begin{equation}
\begin{split}
E_e &= \sum_{k=0}^{L_0-1} \int_{kT_s}^{(k+1)T_s} e^2(t) \,dt \\
&= 4L_0\pi^2 h^2 (M-1)^2\int_{0}^{T_s} \left(q(t)-\frac{t}{2T_s}\right)^2 \, dt
\end{split}
\label{eq:error-energy-full}
\end{equation}
which is basically proportional to the difference between $q(t)$ and the 1REC phase response over a single symbol interval.
\par 
The computation of $E_e$ for partial response signals can be divided into four parts based on the preamble structure as follows. For each part, we denote $E_e$ and $e(t)$ by $E_i$ and $e_i(t)$ respectively. $e_1(t)$ corresponds to the first $L-1$ symbols where CPM modulator does not have any memory of $t<0$. Moreover, we have introduced a shift of $T_l$ for partial response signals. Therefore, $e_1(t)$ during $0\leq t<(L-1)T_s/2$ is expressed as
\begin{equation}
e_1(t)=2\pi h(M-1) \left\{\frac{t}{2T_s}-\sum_{i=0}^{L-2}q\left(t+(\frac{L-1}{2}-i)T_s\right)\right\}
\end{equation} 
which is used to compute $E_1$.
\par 
$E_2$ and $E_3$ correspond to the two intervals, i.e. $T_2$ and $T_3$, where there are transitions in the preamble. Each of these intervals has a duration of $(L-1)T_s$ in which there are at least two symbols with different signs. Due to the symmetry, $E_2=E_3$, we need only to derive $e_2(t)$. At the beginning of $T_2$, the responses of the $L-1$ previous symbols are still in effect, and hence, we must consider $2(L-1)$ symbols, where the first half have a value of $-(M-1)$, and the second half have a value of $M-1$. Additionally, we assume that the time origin is moved to the start of $T_2$ such that $\phi(t)$ is replaced by $\phi_2(t)$. Therefore, $\phi_2(t)$ is expressed as
\begin{equation}
\begin{split}
\phi_2(t) = 2\pi h (M-1) &\left\{\sum_{i=L}^{2(L-1)} q(t+(L-i)T_s)\right. \\
 & \,\, -  \left.\sum_{i=1}^{L-1}  q(t+(L-i)T_s)\right\}.
\end{split}
\end{equation}
We also note that $\phi_2(t)$ is symmetric with respect to $t=(L-1)/2T_s$. Thus, it is sufficient to compare it with the 1REC phase response only for $0\leq t<(L-1)/2T_s$, i.e.,
\begin{equation}
|e_2(t)|=|e_3(t)|=\left|\phi_2(t) + \frac{\pi h (M-1)t}{T_s} + \frac{\pi h(L-1)}{4}\right|.
\end{equation}
\par 
Finally, we need to investigate time intervals where the past $(L-1)$ symbols are similar to the current one. These intervals make up the majority of the preamble with an overall duration of $[L_0-2(L-1)]T_s$ seconds. The absolute value of the error within each such symbol interval is
\begin{equation}
|e(t)| = 2\pi h (M-1)\left|p(t)-\frac{t}{2T_s}-\frac{(L-1)}{4} 
\right|
\end{equation}
where 
\begin{equation}
p(t) =\sum_{i=0}^{(L-1)}q(t+iT_s) 
\end{equation}
for $0\leq t<T_s$. $p(t)$ can be simplified based on $g(t)$, i.e.,
\begin{subequations}
\begin{align}
\label{eq:pt-a}
p(t) \!&= \!\!\!\sum_{i=0}^{(L-1)}\!\! \int_{0}^{iT_s+t}\!\!\!\!\!\! g(\tau)\, d\tau
\! = \!\!\!\sum_{i=0}^{(L-1)} \!\!\int_{0}^{iT_s} \!\!\!\!g(\tau) \, d\tau
+ \!\!\!\sum_{i=0}^{(L-1)} \!\!\int_{iT_s}^{iT_s+t} \!\!\!\!\!\!g(\tau) \, d\tau\\
\label{eq:pt-b}
& =\!\frac{L\!-\!1}{4} +\!\!\! \sum_{i=0}^{(L-1)}\!\! \int_{iT_s}^{iT_s+t} \!\!\!\!\!\! g(\tau)\, d\tau \\
\label{eq:pt-c}
&=\! \frac{L\!-\!1}{4}  + \!\int_{0}^{t} \sum_{i=0}^{L-1} g(\tau+iT_s)\, d\tau
\end{align}
\label{eq:pt}
\end{subequations}
where (\ref{eq:pt-b}) is due to (\ref{eq:qlTs}). It is straightforward to show that the second term in (\ref{eq:pt-c}) is equal to $t/2T_s$ for partial response $L$REC, $L$RC, and Gaussian pulse shapes. Using $L$RC as an example, the integrand in (\ref{eq:pt-c}) can be written as
\begin{equation}
\begin{split}
\sum_{i=0}^{L-1} g(\tau+iT_s) d\tau &= \sum_{i=0}^{L-1}\frac{1}{2LT_s}\left(
1-\cos\frac{2\pi(t+iT_s)}{LT_s}\right)\\
&=\frac{1}{2T_s}-\sum_{i=0}^{L-1} \cos\frac{2\pi(t+iT_s)}{LT_s}=\frac{1}{2T_s} 
\end{split}
\end{equation}
where the last equality holds because the summation of complex points that are uniformly distributed over the unit circle is equal to zero. Hence, $e(t)=0$ for our partial response examples except for $T_2$, $T_3$, and the start of the preamble. This results in 
\begin{equation}
e_a \approx \frac{1}{L_0T_s} \left[\int_{0}^{(L-1)/2T_s} e_1^2(t) dt + 4\int_{0}^{(L-1)/2T_s}e_2^2(t) dt\right] 
\label{eq:err-pr}
\end{equation}
for partial response CPMs of our interest. We note that the integrals in (\ref{eq:err-pr}) are independent of $L_0$. Therefore, the approximation error decreases as $L_0$ increases. We also observe that the approximation error occurs whenever there is a transition in the preamble. Since the optimized preamble has only two transitions, it conveniently limits the approximation error in our approach. On the other hand, if we use (\ref{eq:error-energy-full}) in (\ref{eq:approx-error}), we observe that $e_a$ for full response CPMs is constant with respect to $L_0$. 
\par
\begin{figure}
\centering
\includegraphics[width=0.8\linewidth]{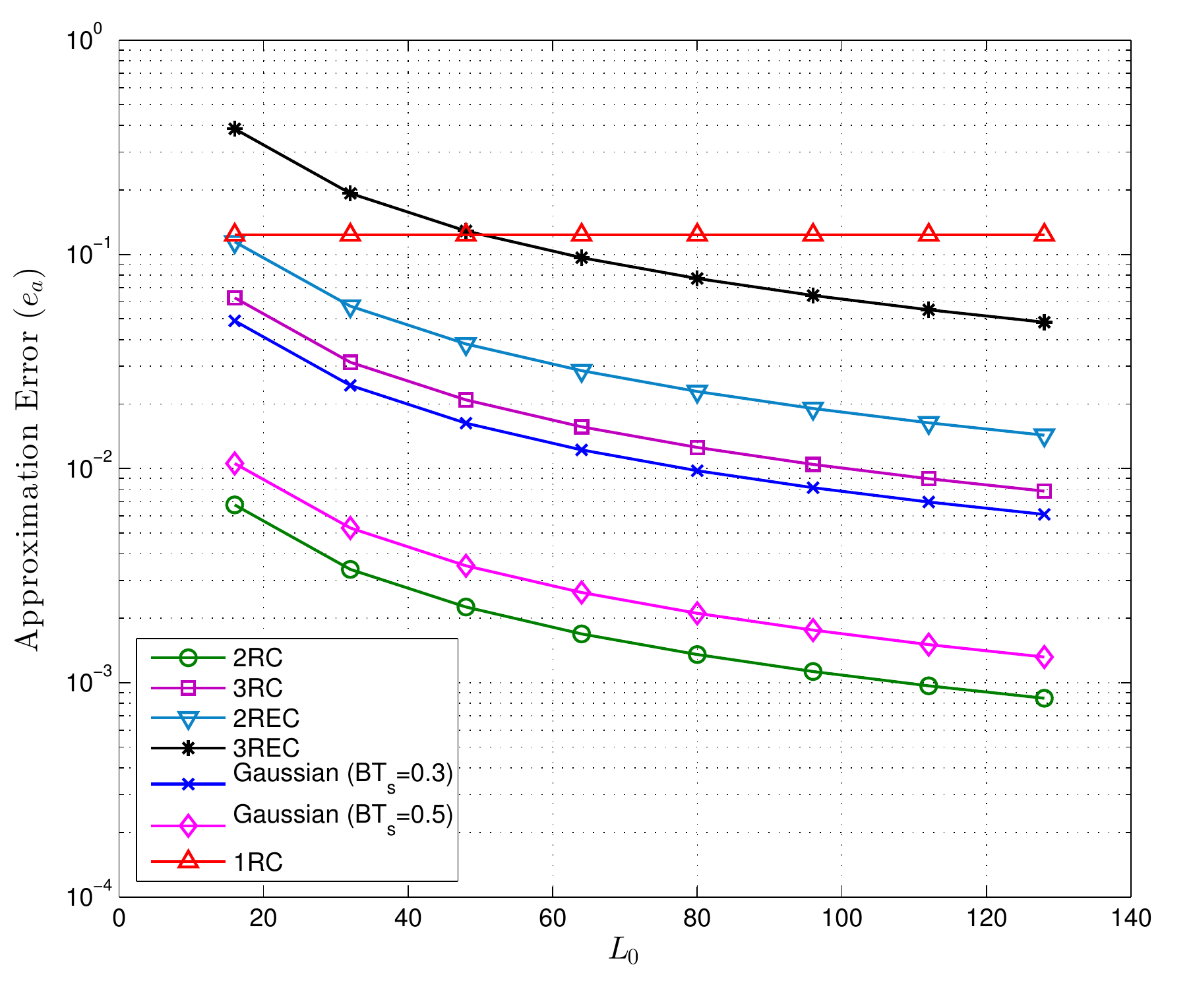}
\caption{The approximation error for different phase responses. These values should be scaled by $h^2(M-1)^2$ for a specific CPM.}
\label{fig:ApproxErr}
\end{figure}

 
We have computed $e_a$ for different CPMs based on the above relations and have plotted them in Fig. \ref{fig:ApproxErr} with respect to $L_0$. We have normalized the curves by $h^2(M-1)^2$ in order to isolate the effect of $L_0$ and $q(t)$ on the approximation error. We observe that longer frequency pulses result in larger approximation errors because the interval for which they exhibit deviations from 1REC is proportional to $L$.  It is also seen that the approximation error for 1RC is much larger than other examples. Its effect on the estimation performance will be seen in the next section.
\subsection{Timing and Carrier Recovery Performance}
In this section, we compute the error variances of frequency offset, carrier phase, and symbol timing for the proposed ML estimation algorithm using simulations. We have considered the three examples of Fig. \ref{fig:SeqResponse} along with MSK, which is a  binary CPM with $h=1/2$ and 1REC frequency pulse. In all examples, the optimum preamble with $L_0=64$ is deployed. In addition to AWGN, we apply $\nu$, $\theta$ and $\varepsilon$ that are uniformly distributed over $[-0.5,0.5)$, $[0,2\pi)$ and $[-0.5,0.5)$ respectively. Additionally, we have considered $N=2$, $K_f=2$, and the Gaussian interpolator (\ref{eq:interpolator}) for the estimation. Our simulations show that both interpolation and zero padding must be present in order to have reliable estimates \cite[Fig. 3]{Hosseini2013b}.
\begin{figure}
        \centering
        \begin{subfigure}[b]{0.5\textwidth}
                \centering
                \includegraphics[width=0.8\textwidth]{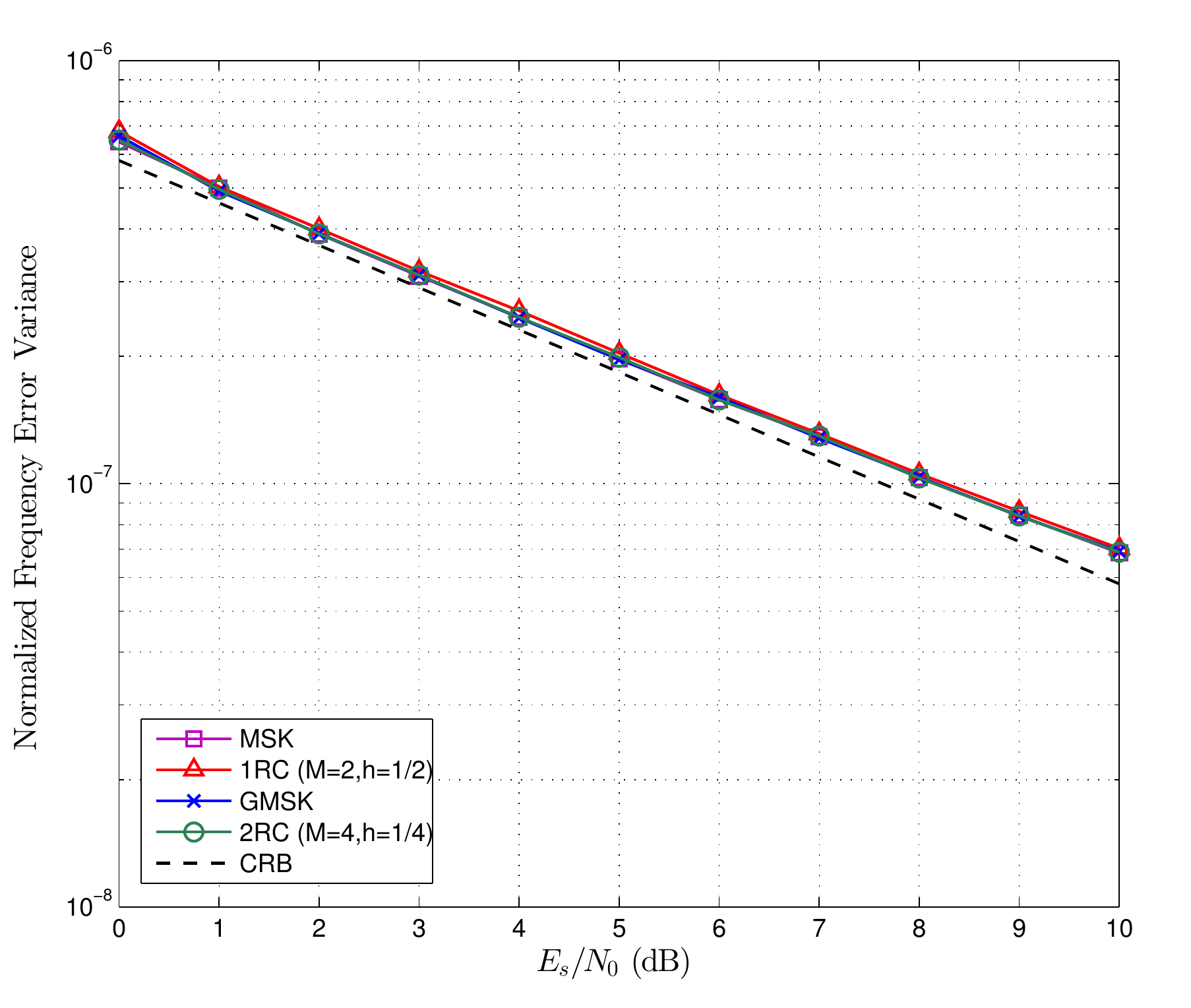}
                \caption{}
                \label{fig:VarF}
        \end{subfigure}%
        
        \begin{subfigure}[b]{0.5\textwidth}
                \centering
                \includegraphics[width=0.8\textwidth]{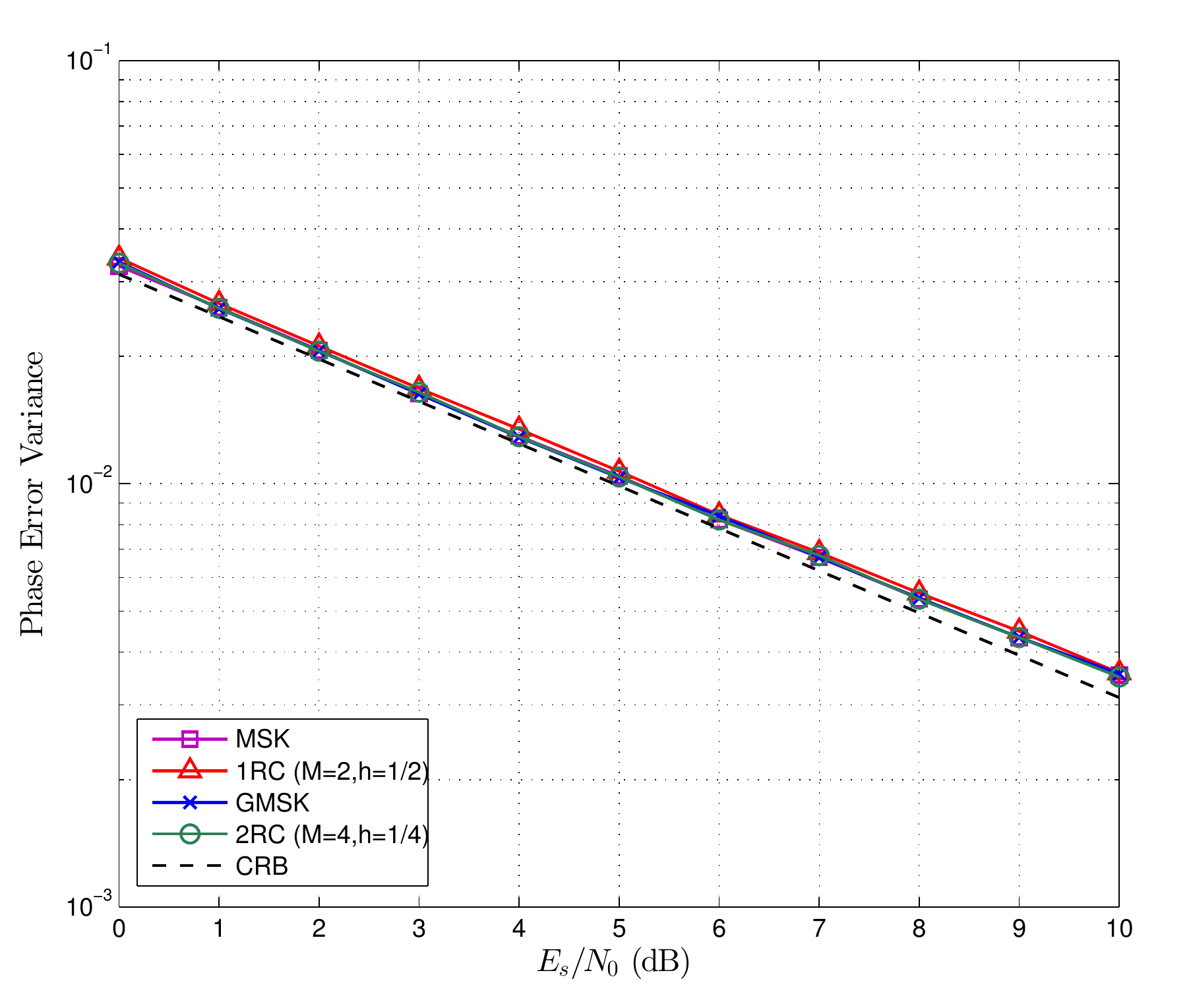}
                \caption{}
                \label{fig:VarPhase}
        \end{subfigure}
        \caption{The error variance of frequency offset (a) and carrier phase (b) estimations for different CPM schemes when $L_0=64$. The frequency is normalized with respect to the symbol rate.}
        \label{fig:VarFP}
\end{figure}
\begin{figure}[t]
\centering
\includegraphics[width=0.8\linewidth]{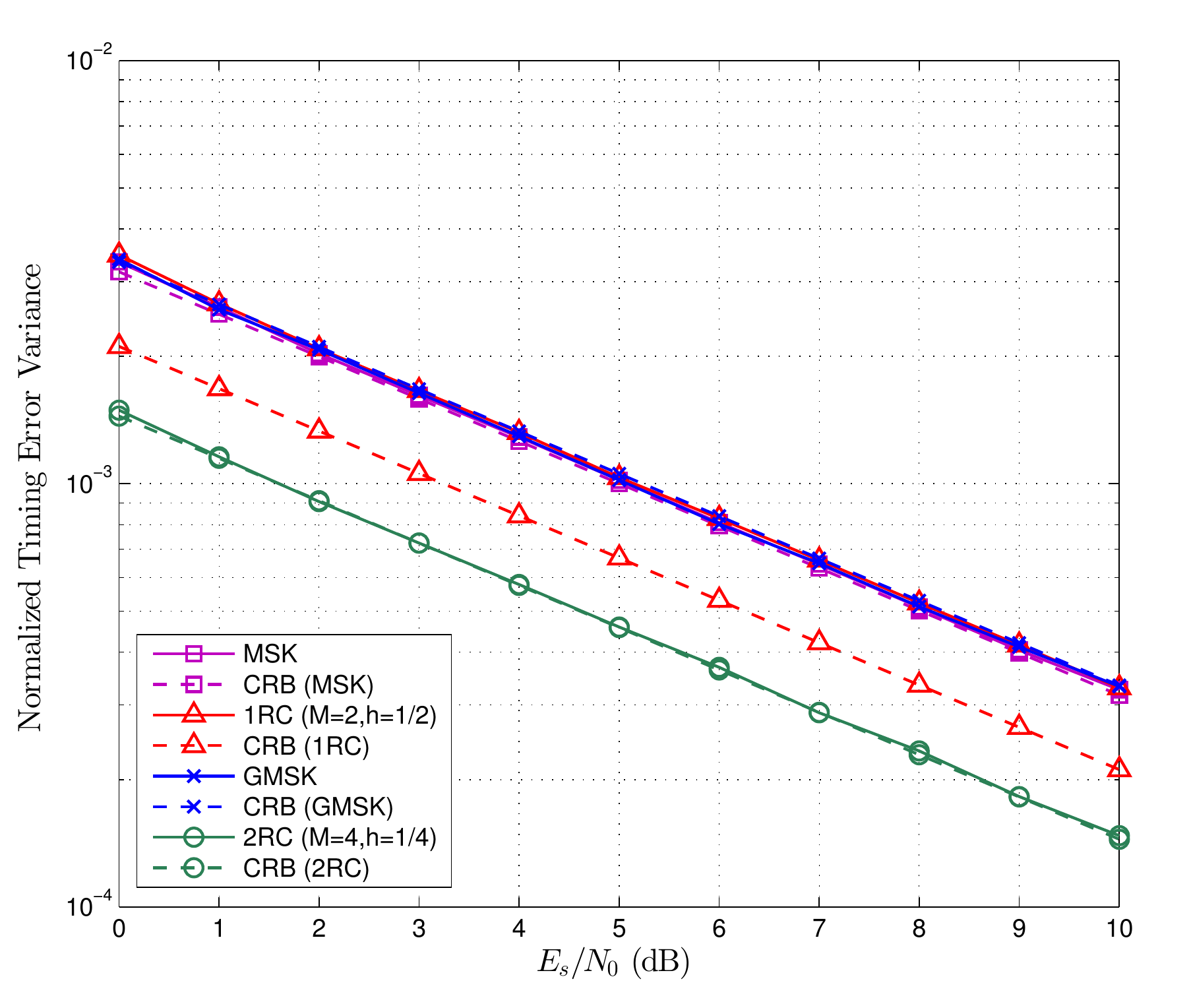}
\caption{The variance of symbol timing estimation for different CPM schemes when $L_0=64$. The symbol timing is normalized with respect to the symbol duration.}
\label{fig:VarTau}
\end{figure}
\par 
The estimation error variances corresponding to the normalized frequency offset and carrier phase are depicted in Figs. \ref{fig:VarFP}\,(a) and \ref{fig:VarFP}\,(b) respectively for different CPM schemes. The frequency estimation plots demonstrate that the proposed estimator performs with almost the same accuracy for all the schemes and is less than 0.5 dB away from the CRB for low to moderate SNRs. As it was shown in \cite{Hosseini2013}, the frequency and phase estimation CRBs for the optimum training sequence are independent of the particular CPM scheme. Hence, only one CRB plot is shown in each figure. Moreover, it is observed that the 1RC scheme performs slightly worse than the other schemes because it has the largest deviations from the 1REC template (refer to Fig. \ref{fig:SeqResponse}). For the remaining schemes, the gap between the error variances and the CRB is mostly due to the FFT precision and can be reduced by increasing $K_f$. This divergence is the beginning of a floor at high SNRs because errors caused by the thermal noise become less significant than the FFT and interpolation precision, which are constant at all SNR.
\par
The normalized timing error variances are plotted in Fig. \ref{fig:VarTau}. It reveals that the proposed estimator reaches the CRB for the majority of the examples. The only exception is again the 1RC scheme as discussed above. For all other examples, the ML estimator attains the lower limit of the CRB despite the visible loss in the frequency estimation. This is because the optimum training sequence decouples timing from frequency in terms of the Fisher information matrix (FIM) \cite{Hosseini2013}, which means that small errors in the frequency estimate do not affect the symbol timing estimate. The optimum training sequence does not decouple frequency and phase, and hence, errors in the frequency estimate leak into the phase estimator, which results in a slight performance degradation that is visible in Fig. \ref{fig:VarFP}\,(b).
\par 
It should be mentioned that the FFT operations will be replaced by simple correlations when $f_d=0$. In such applications, (\ref{eq:lambda1-dis}) and (\ref{eq:lambda2-dis}) are computed for $\tilde{\nu}=0$ without any need to perform the maximization of (\ref{eq:fd-hat}) and the interpolation. This leads to a joint symbol timing and carrier phase estimator, which is efficient yet less complex compared to other DA works such as \cite{Huber1992,Tang2001,Zhao2006a}. This simplicity is a direct result of unique structure of the optimized preamble.
\begin{figure}[t]
\centering
\includegraphics[width=0.8\linewidth]{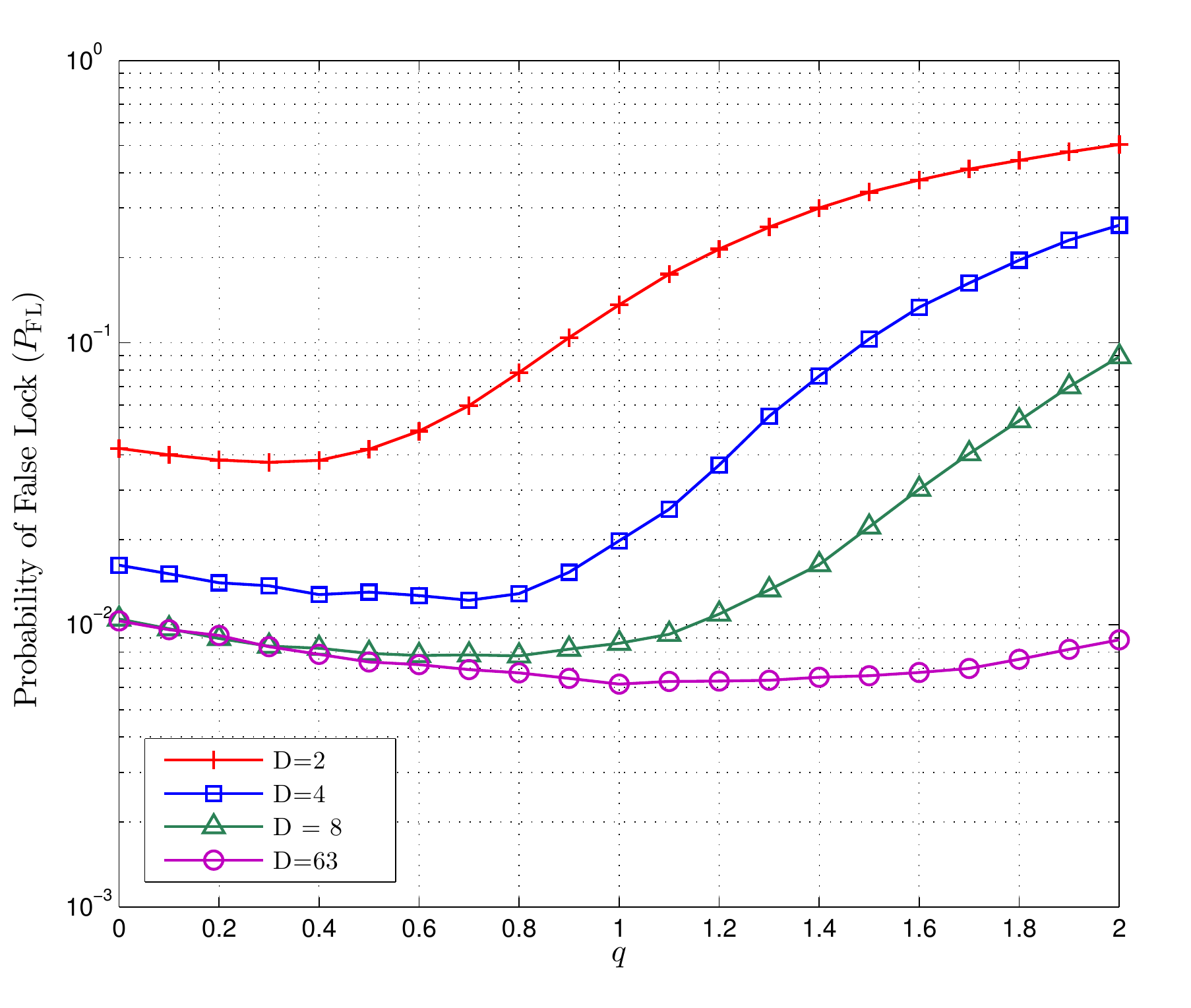}
\caption{The effect of correction term $C(\delta)$ (Equation (\ref{eq:cdelta})) and its exponent $q$ on $P_{\text{FL}}$. GMSK signaling is used when $N_p=64$ and $E_s/N_0=1$ dB.}
\label{fig:PfaQGMSKL64}
\end{figure} 
\subsection{Frame Synchronization Performance}
The performance of the SoS estimation algorithm is characterized by the probability of \textit{false lock}, which is $P_{\text{FL}}=\text{Pr}\{\hat{\delta} \neq \delta \}$. This probability is computed given that the preamble is correctly detected and fully resides within the observation window.
\begin{figure}[t]
\centering
\includegraphics[width=0.8\linewidth]{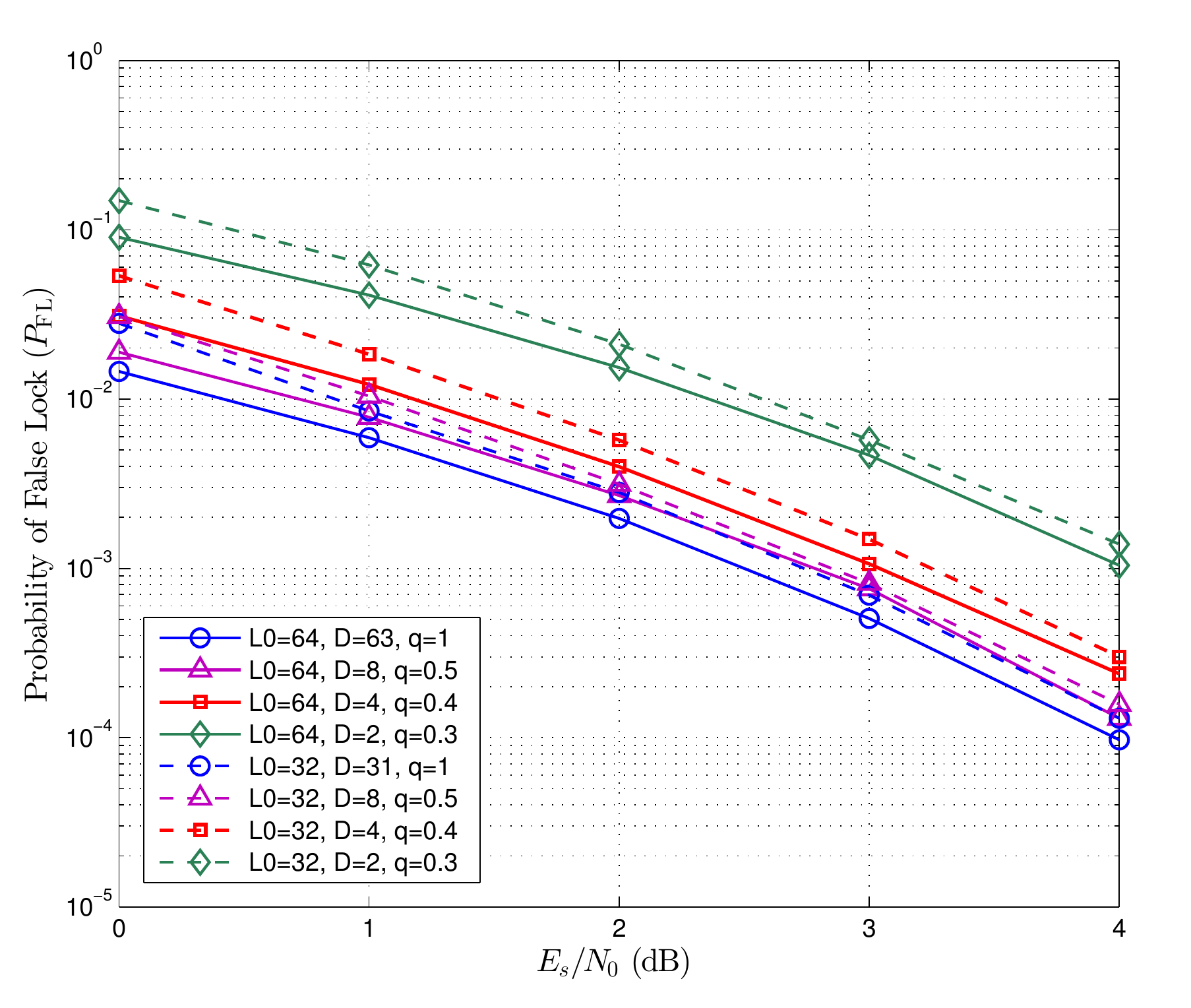}
\caption{The probability of false lock versus SNR for different preamble lengths. The values of $q$ are optimized for each case using simulations at $E_s/N_0=1$ dB. The signal is sampled at $N=1$, which results in $N_p=L_0$.}
\label{fig:PfaGmskSnr}
\end{figure}
\par
The effect of the $C(\delta)$ as a function of $q$ on $P_{\text{FL}}$ is studied in Fig. 
\ref{fig:PfaQGMSKL64} using simulations. The GMSK scheme is used where $L_0=N_p=64$, $N_{w}=96$ and $E_s/N_0 = 1$ dB. Additionally, we have varied $q$ over the range of $[0,2]$ and computed $P_{\text{FL}}$ for several values of $D$. It is observed that the introduction of $C(\delta)$ reduces $P_{FL}$ given that $q$ is carefully selected. We observe that a value of $q=1$ is suitable for $D=63$, while it needs to be decreased for smaller values of $D$. In fact, $C(\delta)$ becomes less important for small values of $D$ such as $D=2$ and can simply be ignored, i.e. $q=0$. Nevertheless, it visibly improves the performance for $D=63$ such that it becomes superior to $D=8$ only in the presence of $C(\delta)$. Our simulations also confirm that the SoS estimator becomes unbiased only for the optimized $q$, which was the main motivation for introduction of $C(\delta)$ as in (\ref{eq:cdelta}).
\par
The SoS estimator's performance with respect to SNR is shown in Fig. \ref{fig:PfaGmskSnr} for two different preamble lengths and multiple values of $D$. Our simulations show that variations of the optimum $q$ with respect to SNR is less than 0.1 for the plotted range. Therefore, we use a fixed $q$ for each plot. We note that the proposed parameter of $D$ allows us to avoid unwieldy complexity of $D=N_p-1$. For instance, choosing $D=4$ results in only a loss of 0.7 dB for $L_0=64$ in comparison with $D=63$. Yet, the computational complexity is reduced by a factor of approximately 16. Another important observation that can be made is that increasing $L_0$ from 32 to 64 yields a gain of only a fraction of dB in terms of the SNR.
\begin{figure}[t]
\centering
\includegraphics[width=0.8\linewidth]{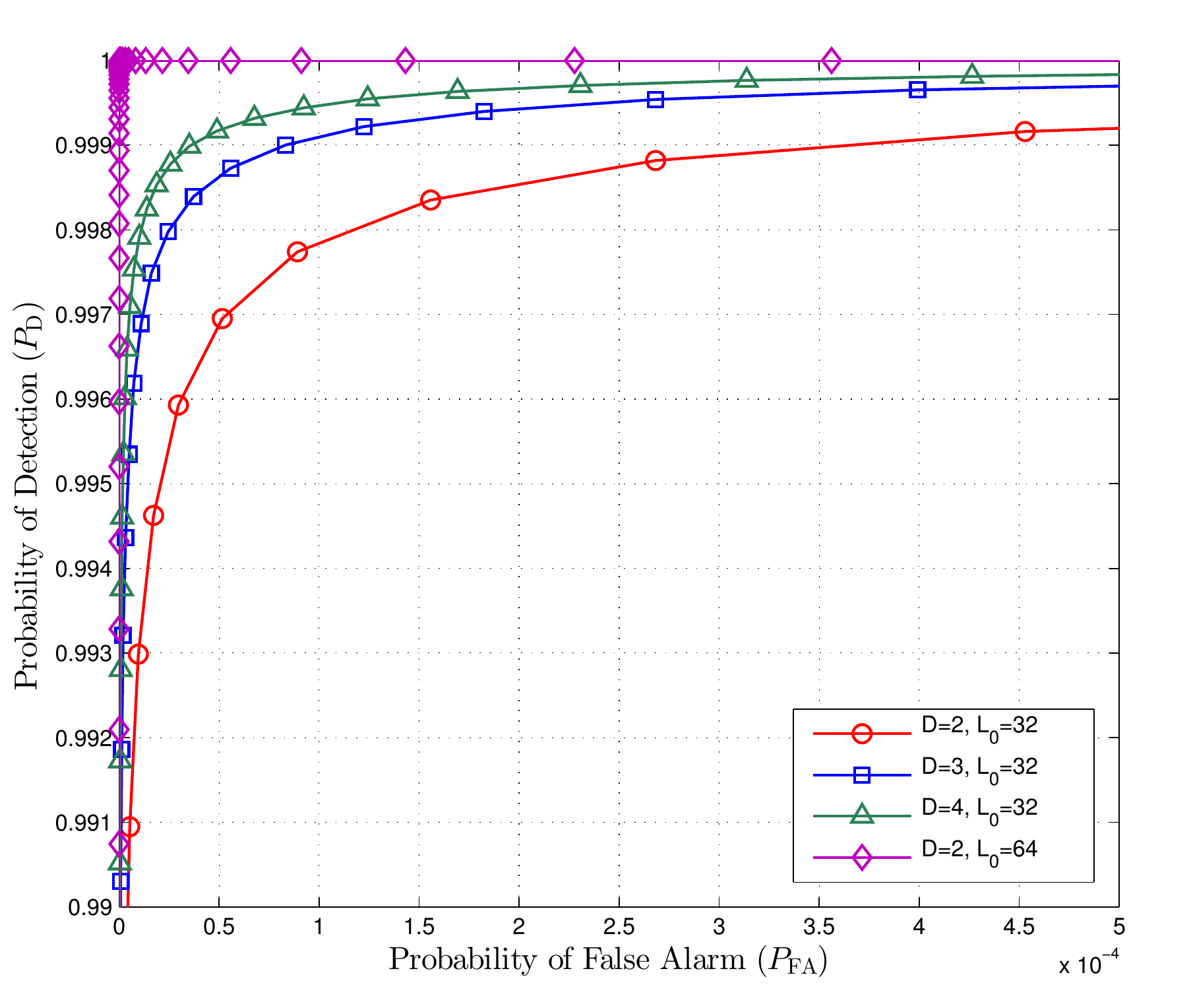}
\caption{Receiver operating characteristics for the proposed detector. The optimum preamble is transmitted over an AWGN channel when $E_s/N_0=1$ dB and GMSK modulation is used.}
\label{fig:RocGMSK}
\end{figure} 
\par 
The performance of the SoS detection algorithm can be examined through the ROC plots. A few examples of the ROC are plotted in Fig. \ref{fig:RocGMSK} where $P_{\text{FA}}$ and $P_{\text{D}}$ are calculated using simulations by varying $\gamma_{D'}$. These ROCs are obtained for the GMSK scheme when $E_s/N_0=1$ dB, and $N=1$ ($N_p=L_0$). It is observed that we are able to attain a very low $P_{\text{FA}}$ at low SNR even with a relatively short preamble of $L_0=32$. It is also seen the improvement becomes less significant when $D'$ is changed from 4 to 8. Therefore, a small value of $D'$ looks sufficient to achieve a $P_{\text{D}}$ that is close to the full-complexity detector, i.e. $D'=N_p-1$. This is similar to the performance improvement of the SoS estimation algorithm versus $D$  (Figs. \ref{fig:PfaQGMSKL64} and \ref{fig:PfaGmskSnr}). On the other hand, the performance is improved substantially when $L_0=64$. For instance, $P_{\text{D}}=1-5\times 10^{-7}\approx 1$ and $P_{\text{FA}}=4.86\times 10^{-6}$ for $\gamma_2=40$. Comparing these metrics with $P_{\text{FL}}$ in Fig. \ref{fig:PfaGmskSnr} reveals that the performance of the frame synchronization algorithm is limited by the false locks rather than false alarms or missed detections.   
\subsection{BER Performance}
\begin{figure}
        \centering
        \begin{subfigure}[b]{0.5\textwidth}
                \centering
                \includegraphics[width=0.8\textwidth]{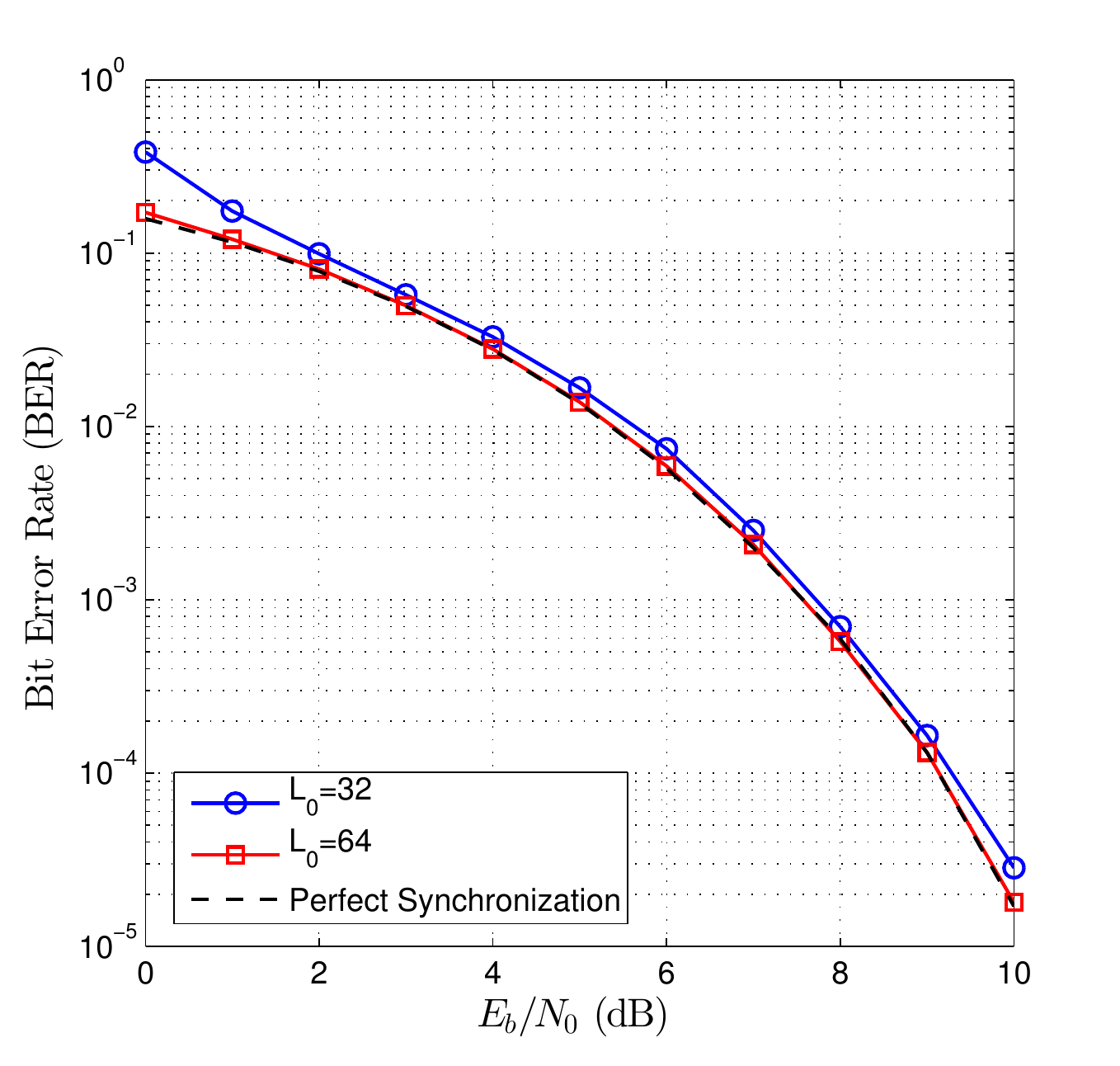}
                \caption{GMSK $(BT_s=0.3)$}
                \label{fig:GmskBer}
        \end{subfigure}%
        
        \begin{subfigure}[b]{0.5\textwidth}
                \centering
                \includegraphics[width=0.8\textwidth]{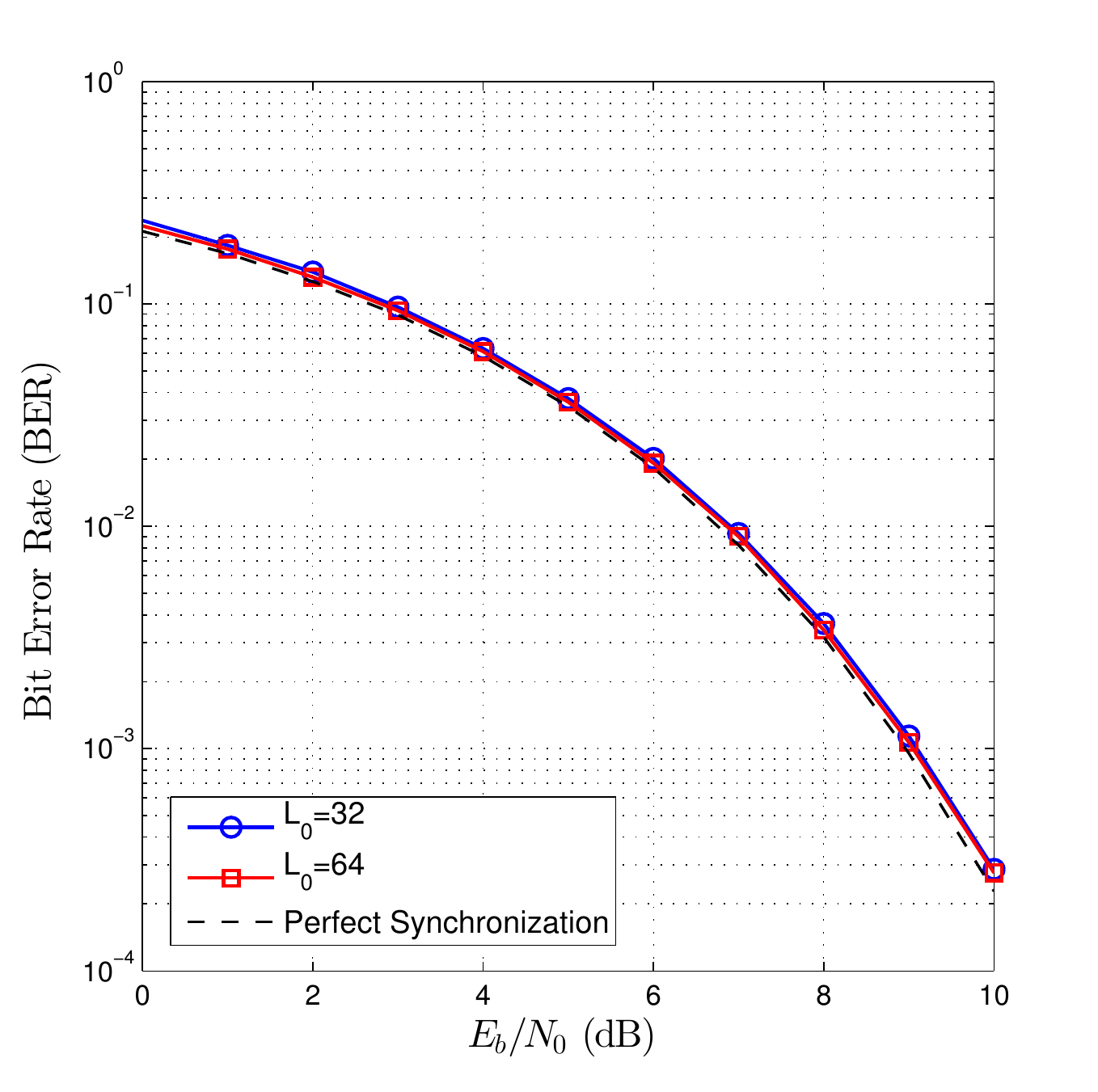}
                \caption{2RC $(h=1/4, M=4)$}
                \label{fig:2RcBer}
        \end{subfigure}
        \caption{BER for the burst-mode CPM receiver. $L_0$ is the preamble length in terms of data symbols.}\label{fig:BERs}
\end{figure}
In this section, we evaluate the overall BER performance of the proposed synchronization scheme, including SoS detection, SoS estimation, and timing and carrier recovery, using simulations. We have considered two examples of GMSK and 4-ary, $h=1/4$ CPM with 2RC frequency pulse. Each burst consists of a preamble of $L_0$ symbols, a UW of 64 random but known bits and 4096 information bits. The UW is used in order to adjust the beginning of each burst by correlating it with the demodulated bits. In our simulations, the transmitter sends individual bursts that are preceded by a fixed but unknown amount of guard time. The AWGN is then added to the waveform along with random frequency and phase offsets. The received signal is sampled at $N=2$ samples per symbol. The MLSD CPM demodulator is designed according to \cite{Anderson1986}, which uses the Viterbi algorithm (VA). We have also employed a decision-directed (DD) phase and timing tracking loop \cite{Morelli1997} in which phase and timing error signals are generated according the decisions made inside the VA. The phase tracking loop is essential because even very small residual frequency offsets, after the DA synchronization, result in large phase rotations as the burst is being demodulated. The phase and timing loop bandwidths are both set to $10^{-3}/T_s$. Finally, we have set $D'=D=4$ and $N_w=2NL_0$ for the frame synchronization.
\par
The BER performance of the burst-mode receiver for the GMSK scheme with two different preamble lengths is depicted in Fig. \ref{fig:BERs}\,(a). It is observed that the receiver operates within less than 0.1 dB of the ideal synchronization for $L_0=64$ over the whole range of $E_b/N_0$. However, there is a substantial BER degradation at the low SNR region for the short preamble of $L_0=32$. Our simulation results show that this is mainly due to the SoS false locks that are more likely to happen at low SNRs and short preamble lengths. False locks reduce the accuracy of the timing and carrier recovery algorithm, which impact the BER. At higher SNRs, there is about 0.2 dB gap that is caused by estimation errors, which are increased when $L_0$ is reduced.
\par 
The BER performance for the 2RC scheme is reported in Fig. \ref{fig:BERs}\,(b). Similar to GMSK, $L_0=64$ performs almost ideal and within about 0.1 dB of perfect synchronization. However, the preamble of $L_0=32$ shows slightly different behavior than that of GMSK, where no BER degradation at low SNRs is visible. This is because $E_s=2E_b$ for the 4-ary scheme and both Figs. \ref{fig:BERs}\,(a) and \ref{fig:BERs}\,(b) are expressed in $E_b/N_0$. In other words, $E_s/N_0$ for GMSK is 3 dB less than for the 2RC, and hence, $P_{\text{FL}}$ becomes larger. In fact, 2RC with $L_0=32$ should be compared to GMSK with $L_0=64$ in order to have a fair comparison where both preambles contain 64 bits. We also note there is no visible difference between the two preambles in terms of the BER, and hence, $L_0=32$ is an adequate length in practice. Finally, this scheme, i.e. non-binary and partial response, is known to be prone to false locks when DD timing estimation algorithms such as \cite{Huber1992} or \cite{Morelli1997} are implemented. Here, we showed that our proposed DA algorithm with a short preamble can be another method to solve the false lock problem while it significantly reduces the acquisition time.           
\section{Conclusion}
In this paper, we addressed the synchronization problem for CPM signals in burst-mode transmissions. Thanks to the unique structure of the optimized synchronization preamble, we developed a DA ML algorithm, which jointly estimates the frequency offset, carrier phase and symbol timing. The proposed algorithm, which is implemented in a feedforward manner, estimates the frequency offset via two FFT operations. Once the frequency estimate is available, the carrier phase and symbol timing are easily computed via simple closed-form expressions. Our method can be applied to the whole range of CPM signals. The computed MSEs demonstrate that its performance is within 0.5 dB of the CRB for all three synchronization parameters for various examples. Moreover, it operates at frequency offsets as large as half of the sampling frequency without sacrificing the estimation accuracy.
\par
In the second part of this paper, we addressed the frame synchronization issue in burst-mode CPM transmissions using ML principles. We developed a simple test for detection of the SoS after which the exact location of the SoS is estimated via a one-dimensional search. We numerically computed the ROCs for the SoS detector along with the false lock probabilities for the SoS estimator. The frame synchronization allowed us to implement a realistic burst-mode CPM receiver. The simulated BER curves demonstrated an almost ideal performance for preambles as short as 64 bits and SNRs as low as 0 dB.  

%
\appendix
\section*{Derivation of $T_l$}
We start by assuming transmission of $K$ ``$M-1$'' symbols when the phase response length is $L$. The CPM phase at $t=KT_s$ when $K>L$ can be written as
\begin{equation}
\begin{split}
\phi(KT_s) &= 2\pi h \sum_{i=0}^{K-1} (M-1)q(KT_s-iT_s) \\
&=\pi h (M-1)(K-L+1)+\! 2\pi h (M-1) \sum_{l=1}^{L-1}\!q(lT_s)
\end{split}
\label{eq:PhiKTs}
\end{equation}
where the second equality holds since $q(mT_s)=1/2$ for $m\geq L$. Without loss of generality we assume $L$ is odd. Additionally, we consider frequency pulses which have even symmetry around $LT_s/2$. Therefore, the second term on the right-hand side of (\ref{eq:PhiKTs}) can be expressed as
\begin{equation}
\begin{split}
\sum_{l=1}^{L-1}q(lT_s) &= \sum_{k=1}^{(L-1)/2}q(kT_s)+q((L-k)T_s)  \\
&= \sum_{k=1}^{(L-1)/2} \int_{0}^{(L/2)T_s} g(t)dt -\int_{kT_s}^{(L/2)T_s} g(t) dt \\&\qquad+ \int_{0}^{(L/2)T_s} g(t)dt  + \int_{(L/2)T_s}^{(L-k)T_s} g(t) dt  \\
&= \sum_{k=1}^{(L-1)/2} \frac{1}{2} = \frac{L-1}{4}.
\end{split}
\label{eq:qlTs}
\end{equation}
The last equality is true due to the following equalities for symmetric $g(t)$,
\begin{align}
&\int_{0}^{(L/2)T_s} g(t)dt = \frac{1}{2} \int_{0}^{LT_s} g(t)dt = \frac{1}{4} \\
&\int_{kT_s}^{(L/2)T_s} g(t) dt = \int_{(L/2)T_s}^{(L-k)T_s} g(t) dt.
\end{align} 
where $k<L/2$. Thus, (\ref{eq:PhiKTs}) is simplified to
\begin{equation}
\phi(KT_s) = \pi h (M-1)[K-\frac{L-1}{2}].
\label{eq:PhiKTsSimp}
\end{equation}
It can be shown that the above results hold for even values of $L$ as well. It is observed that the signal phase in (\ref{eq:PhiKTsSimp}) is equal to the phase of a CPM signal with 1REC pulse shape, same $h$ and data sequence at $t=(K-\frac{L-1}{2})T_s$. The latter signal is basically the approximated phase response, and hence,
\begin{equation}
T_l = KT_s - (K-\frac{L-1}{2})T_s = \frac{L-1}{2}T_s.
\label{eq:Tl}
\end{equation}
\bibliographystyle{ieeetr}
\bibliography{library}
\end{document}